\documentclass[slac_one]{revtex4}
\usepackage{graphicx}
\usepackage{fancyhdr}
\usepackage{units}

\newcommand{\ee}{e^{+}e^{-}}

\newcommand{\jp}{J/\psi}
\newcommand{\ec}{\eta_{\,c}}
\newcommand{\hc}{h_{c}}

\newcommand{\etap}{\eta^{\prime}}
\newcommand{\etac}{\eta_{\, c}}
\newcommand{\psip}{\psi^{\prime}}

\newcommand{\pipi}{\pi^{+}\pi^{-}}
\newcommand{\piz}{\pi^{0}}

\newcommand{\DDbar}{D\bar{D}}
\newcommand{\ccbar}{c\bar{c}}
\newcommand{\bbbar}{b\bar{b}}

\newcommand{\ppbar}{p\bar{p}}

\newcommand{\rt}{\rightarrow}

\newcommand{\etal}{\em et al.}
\newcommand{\jpsi}{J/\psi}
\newcommand{\azero}{a_{\, 0}(980)}

\newcommand{\fzero}{f_{\,0}(980)}

\newcommand{\BR}{{\cal B}}
\newcommand{\etacp}{\eta_c(2S)}
\newcommand{\KsKpi}{K^0_SK^\pm\pi^\mp}
\newcommand{\kk}{K^+K^-}        
\newcommand{\kkpiz}{\kk\piz}          

\newcommand{\gm}{\gamma}
\newcommand{\gmsm}{\gamma_{\unit{sm}}}
\newcommand{\br}[1]{\mathcal{B}(#1)}        
\newcommand{\chicj}{\chi_{cJ}}
\newcommand{\gevcc}{\,\unit{GeV}/\unit{c}^2}
\newcommand{\mm}{\mu^+\mu^-}

\begin{document}

\title{Recent Results from BESIII} 

%

\author{Guangshun Huang (Representing the BESIII Collaboration)}
\affiliation{Department of Modern Physics,
        University of Science and Technology of China,
        96 JinZhai Road, Hefei, Anhui, 230026, P.R.China}

\begin{abstract}

BESIII is a new state-of-the-art 4$\pi$ detector at the upgraded
BEPCII two-ring $\ee$ collider at the Institute of High Energy
Physics in Beijing, China.  It has been in operation since 2008,
and has collected the world's largest data samples of $\jpsi$, 
$\psip$ and $\psi(3770)$ decays, as well as $\tau$ mass scan
and low energy points for $R$ measurement. These data are being 
used to make a variety of interesting and unique studies
of light-hadron spectroscopy, precision charmonium physics,
high-statistics measurements of $D$ meson decays, $\tau$ mass
measurement and $R$ measurement.  Results summarized in
this report include observations of a subthreshold $p\bar{p}$
resonance in $\jp\to\gamma p\bar{p}$, a large isospin-violation 
in $\eta(1405)\rt\piz\fzero$ decays, a near-threshold 
enhancement in $\jpsi\rt\gamma\omega\phi$, and a M1 transition 
$\psip\rt\gamma\eta_c(2S)$; the $\rho\pi$ puzzle in $\jpsi$ and 
$\psip$ decays; some recent precision measurements of $\ec$ and 
$\hc$ lineshapes; and preliminary results of the $D$ meson 
(semi-)leptonic decays and the $\tau$ mass measurement.

\end{abstract}

\maketitle

\thispagestyle{fancy}


\section{Introduction}

The BES experimental program dates back to late 1989 when 
operation of the Beijing Electron Positron Collider (BEPC) 
and the Beijing Electron Spectrometer (BES) first started.
BEPC was a single-ring $\ee$ collider that operated in the
$\tau$-charm threshold energy region between 2.0~and~5.0~GeV with
a luminosity of $\sim 10^{31}$cm$^{-2}$s$^{-1}$.  Among the early
successes included a precise measurement of the mass of the $\tau$ lepton~\cite{tau-mass}
that not only improved on the precision of previous measurements by an order-of-magnitude,
but also showed that the existing world avearge value was high by about two standard
deviations.   
In the late 1990s, the BES detector was upgraded to the BESII detector
and this produced another key result that was the precise measurement 
of the total cross section for $\ee$ annihilation into hadrons ($R$ value)
over the accessible center of mass (c.m.) energy range~\cite{bes_R, bes_R2}.
The precision of these measurements lead to a substantially improved 
evaulation of the electromagnetic coupling constant extrapolated to 
the $Z$-boson mass peak, $\alpha_{QED}(M^2_Z)$, which resulted in a 
significant $\sim$30\% increase in the Standard Model (SM) predicted 
value for the Higgs' boson mass~\cite{Higgs-mass}.
BESII also discovered a number of new hadron states, including the 
$\sigma$~\cite{bes_sigma} and $\kappa$~\cite{bes_kappa} scalar 
resonances and a still-unexplained subthreshold $\ppbar$ resonance 
produced in radiative $\jpsi\rt\gamma\ppbar$ decays~\cite{bes_x1860}.

Between 2005 and 2008, BEPC was replaced by BEPCII, a two-ring $\ee$ collider
with a hundred-fold increase in luminosity, and  the BESII detector was completely removed and
replaced by BESIII, a state-of-the-art detector built around a 1~T superconducting
solenoid that contains a cylindrical drift chamber, a double-layer barrel of
scintillation counters for time-of-flight measurements, and a nearly $4\pi$ array of 
6240 CsI(Tl) crystals for electromagnetic calorimetry.  The magnet's iron flux-return  
yoke is instrumented with a nine-layer RPC muon identification system.
BEPCII operations started in summer 2008 and since then
the luminosity has been continuously improving; now it is
$\sim 6.5\times 10^{32}$cm$^{-2}$s$^{-1}$, quite near the  $10^{33}$ design value.
The BESIII detector performance is excellent: the charged particle momentum
resolution is $\delta p/p\simeq 0.5$\%; the $\gamma$ energy resolution is 
2.5\% at $E_{\gamma}=1$~GeV; the 6\% resolution  $dE/dx$  measurements in the drift chamber 
plus the  $\sim$80~ps resolution time-of-flight
measurements is sufficient
to identify charged particles over the entire momentum range of interest.

The BESIII experimental program addresses issues in light hadron physics,
charmonium spectroscopy and decays, $D$ and $D_s$ meson decays, and 
numerous topics in QCD and $\tau$-lepton physics.  To date, BESIII
has accumulated data samples corresponding to 225M (plus $\sim$1.0B in 2012) 
$\jpsi$ decays, 106M (+ $\sim$0.4B) $\psip$ decays, 2.9~fb$^{-1}$ at the peak 
of the $\psi(3770)$ resonance, which decays to $\DDbar$ meson pairs 
nearly 100\% of the time, 24 pb$^{-1}$ around $\tau$-pair threshold,
and 10$k$ hadronic events at each of 4 low energies.
These are all world's-largest data samples at these c.m. energies
and the $\jpsi$ sample is the first ever to be collected in
a high quality detector like BESIII.
In this talk I review some recent results that
have been generated from these data samples.

\section{Light hadron physics}

\subsection{The subthreshold $\ppbar$ resonance seen in $\jpsi \rt \gamma \ppbar$}

As mentioned above in the introduction, BESII reported a peculiar
mass-threshold enhancement in the $\ppbar$ invariant mass distribution
in radiative $\jpsi\rt\gamma\ppbar$ decays~\cite{bes_x1860}.  The
shape of this low-mass peak cannot be reproduced by any of the 
commonly used parameterizations for final state interactions (FSI)
between the final-state $p$ and $\bar{p}$. 

\begin{figure}
\begin{center}
\includegraphics[width=0.8\textwidth]{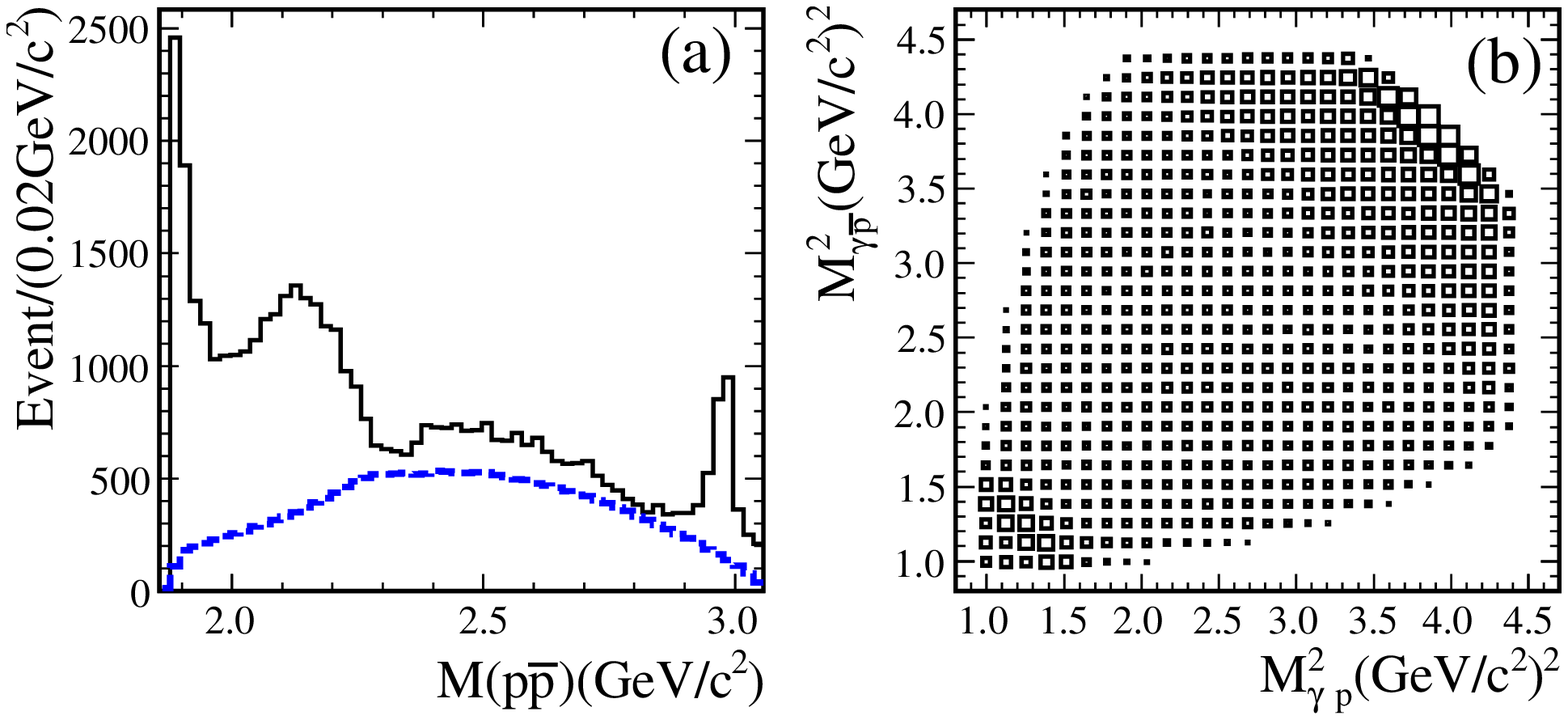}
\caption{{\bf a)} The $M(\ppbar)$ distribution from $\jpsi\rt\gamma\ppbar$ 
decays.  The dashed curve is background from $\jpsi\rt\piz\ppbar$,
where one of the photons from the $\piz\rt\gamma\gamma$ decay has low energy
and is undetected.  The narrow peak on the right is from 
$\jpsi\rt\gamma\etac$, $\etac\rt\ppbar$.
{\bf b)}  The $M^2(\gamma\bar{p})$ (vertical) {\it vs.} $M^2(\gamma p)$
Dalitz plot for the same data sample.  The diagonal band at the upper right
is produced by the $\ppbar$ mass-threshold enhancement; the band at the
lower left is due to the $\etac$.
}
\label{fig:x1860}
\end{center}
\end{figure}

The $\ppbar$ invariant mass
distribution for $\jpsi\rt\gamma\ppbar$ decays in the 225M event BESIII $\jpsi$ data sample
is shown in Fig.~\ref{fig:x1860}a, where the threshold enhancement
is quite prominent~\cite{bes3_x1860}.   A Dalitz plot for these events
is shown in Fig.~\ref{fig:x1860}b.  A partial-wave-analysis (PWA)
applied to these data determined that the $J^{PC}$ of the near-threshold
structure is $0^{-+}$.   A fit using a sub-threshold
resonance shape modified by the Julich FSI effects~\cite{julich_fsi}
yields a mass of $M=1832^{+32}_{-26}$~MeV
and a 90\% CL upper limit on the width of $\Gamma<79$~MeV.

\subsection{Isospin violations in $\eta(1405)$ decays}

\begin{figure}
\mbox{
  \includegraphics[width=0.48\textwidth]{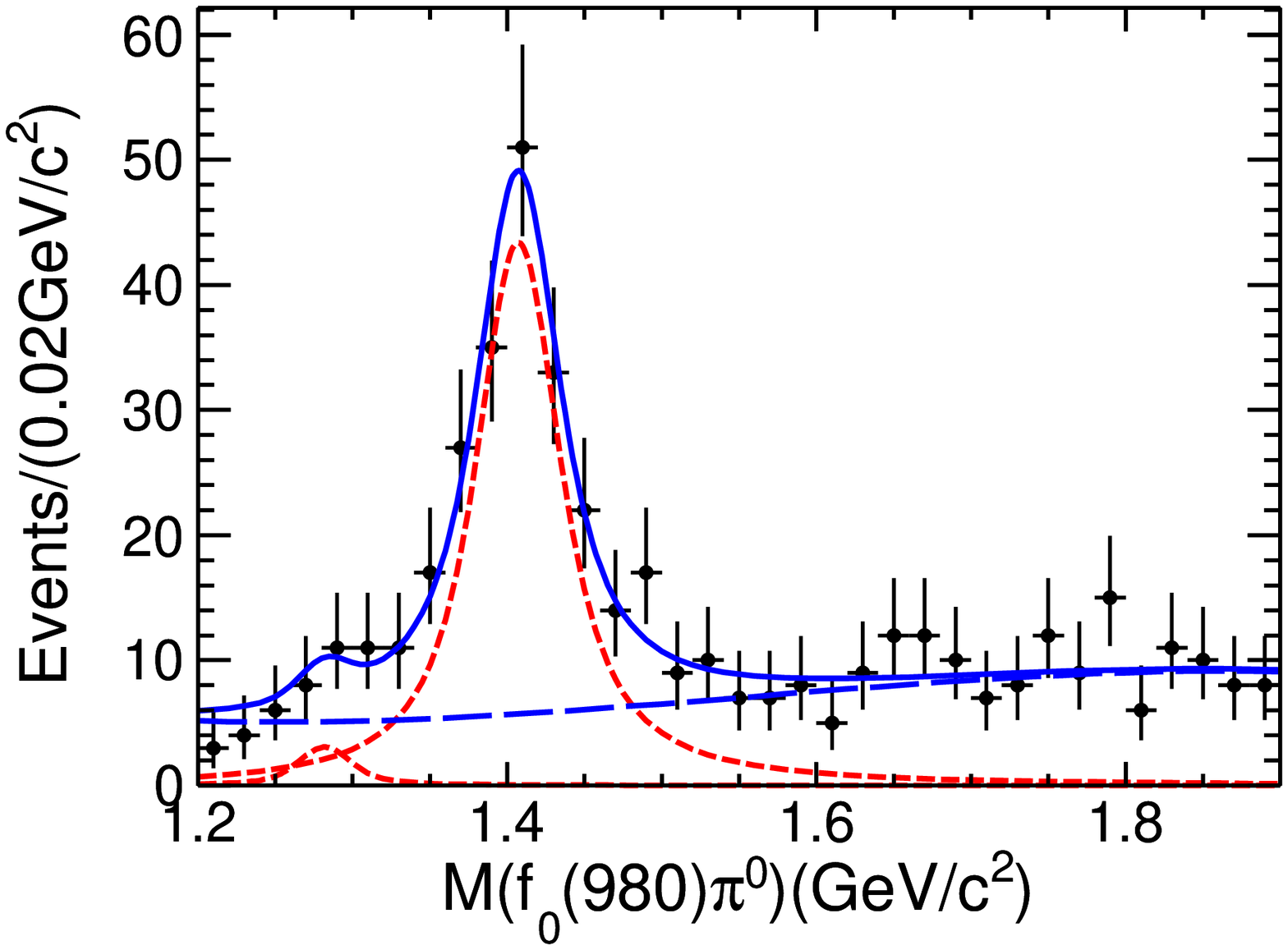}
}
\mbox{
  \includegraphics[width=0.48\textwidth]{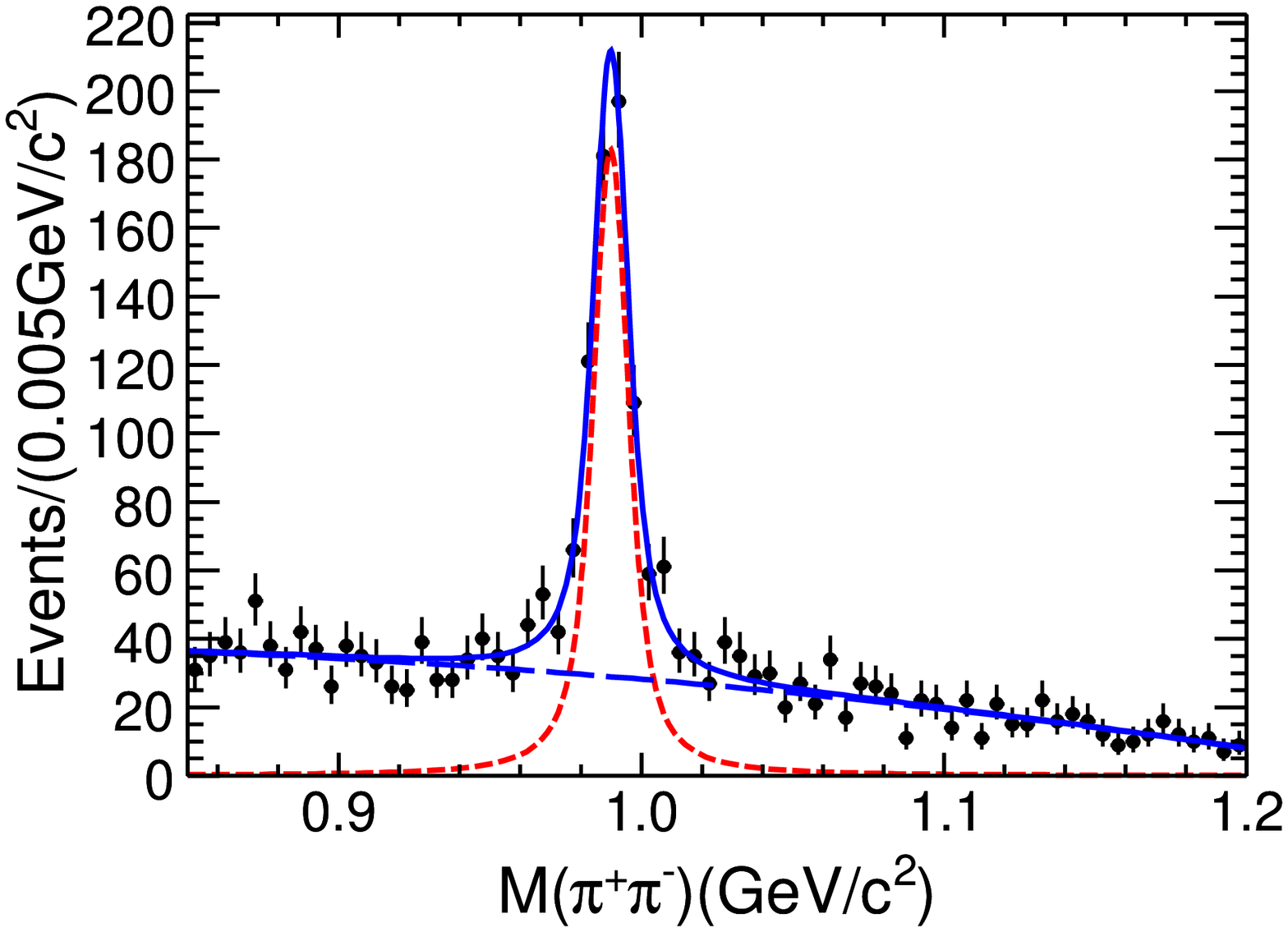}
}
\caption{{\bf Left:} The $\piz f_0$ mass distribution from $\jpsi\rt \gamma \piz f_0$,
decays where $f_0\rt\piz\piz$ (a $\jpsi\rt 7\gamma$ final state!).
{\bf Right:} The $\pipi$ mass distribution from 
$\eta(1405)\rt\piz f_0$ decays where $f_0\rt\pipi$.} 
\label{fig:pi0f0}
\end{figure}

BESIII examined the $\piz f_0$ invariant mass distribution 
produced in radiative $\jpsi\rt \gamma\piz f_0$
decays for both the $f_0\rt \pipi$ and $f_0\rt\piz\piz$ decay
modes~\cite{bes3_eta1405}.  In the distribution for $f_0\rt \piz\piz$ decays, shown in
the left panel of Fig.~\ref{fig:pi0f0} (the $f_{\,0}\rt\pipi$ channel
looks similar), the dominant feature is a pronounced
peak near $M(\piz f_0)=1405$~MeV; helicity analyses
indicate that this peak has $J^p=0^-$, which leads to its
identification as the $\eta(1405)$ resonance.

The decay $\eta(1405)\rt\piz f_0$ violates isospin.  In this
case the observed isospin violation is quite large:
\begin{equation}
\frac{Bf(\eta(1405)\rt\piz \fzero\rt\piz\pipi)}{Bf(\eta(1405)\rt \piz\azero\rt\piz\piz\eta)}
= (17.9\pm 4.2)\%,
\end{equation}
which is an order-of-magnitude larger than is typical for isospin violations. (For
example, BESIII also reports that the isospin violating
$Bf(\etap\rt\pipi\piz)$ is ($0.9 \pm 0.1$)\% of the isospin
conserving $Bf(\etap\rt\pipi\eta)$~\cite{bes3_eta1405}.)

A striking feature 
of these decays is the lineshape of the $f_{\, 0}\rt\pi\pi$ decays,
shown for the $f_{\, 0}\rt\pipi$ channel in the right panel of
Fig.~\ref{fig:pi0f0}, where it can be seen that the
$f_{\, 0}$ peak position is significantly above its nominal $980$~MeV value, and
its width is much narrower than its nominal value of $\sim$100~MeV.
The fitted mass is $M=989.9\pm0.4$~MeV, midway between 
$2m_{K^+}$ and $2m_{K^0}$, and the fitted width is $\Gamma = 9.5\pm 1.1$~MeV,
consistent with the $2m_{K^0} - 2m_{K^+} = 7.8$~MeV mass threshold
difference.

\begin{figure}
\begin{center}
\includegraphics[width=0.5\textwidth]{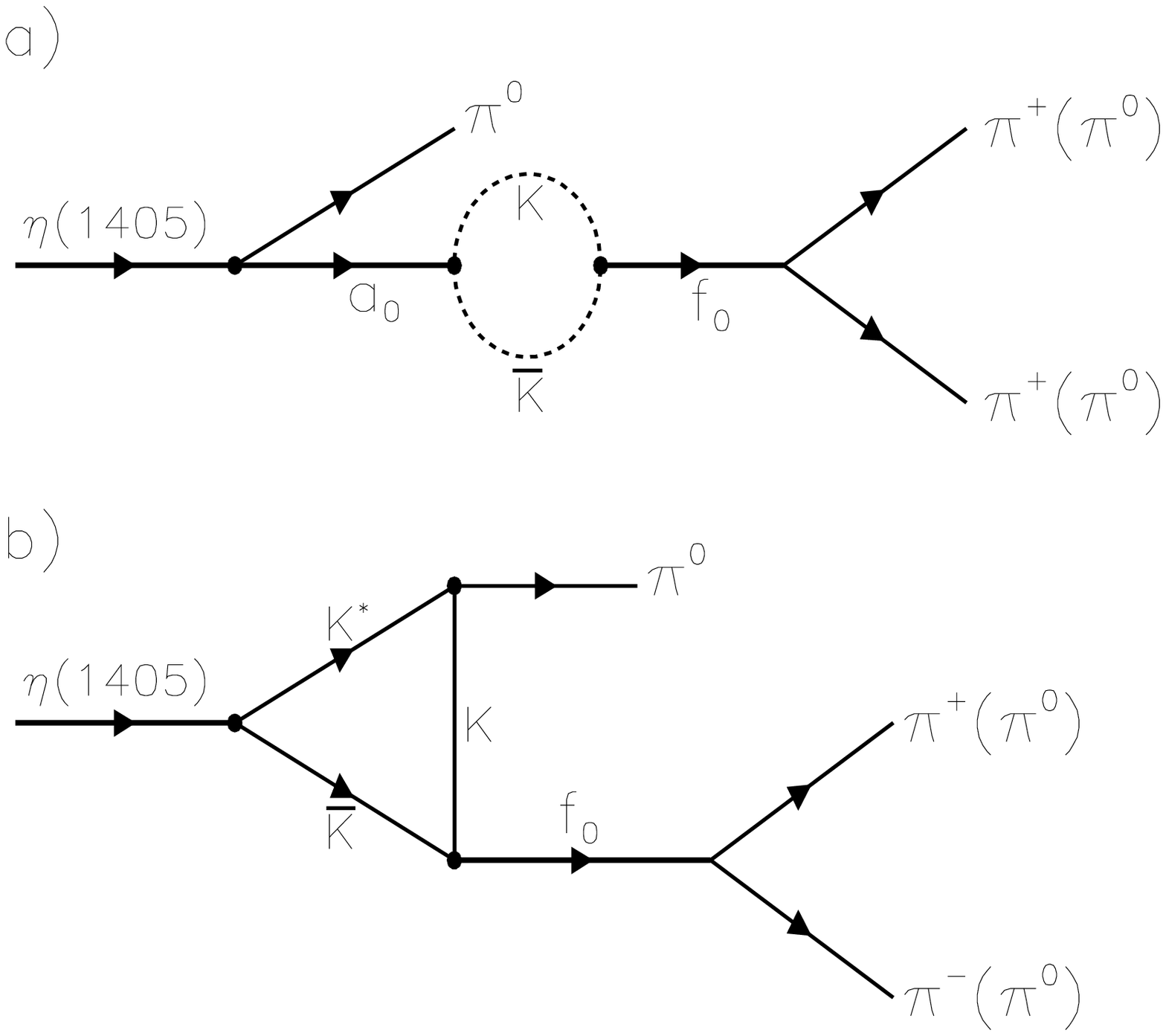}
\caption{{\bf a)} The leading diagram for $\eta(1405)\rt\piz f_0$
via $\azero$-$\fzero$ mixing.  
{\bf b)} The triangle anomaly
diagram in $\eta(1405)\rt \piz\fzero$ decay~\cite{zhao}.
} 
\label{fig:eta1405_triangle}
\end{center}
\end{figure}

Possible processes that mediate $\eta(1405)\rt\piz f_0$ are shown in
Fig.~\ref{fig:eta1405_triangle}.  As we have seen above, the 
$\azero\rt\fzero$ process (Fig.~\ref{fig:eta1405_triangle}a)
is at or below the percent level, and
is too small to account for the large isospin violation that
is observed.  Wu and collaborators~\cite{zhao} suggest that the 
triangle anomaly diagram shown in Fig.~\ref{fig:eta1405_triangle}b
could be large enough to account for the data.  In this case, both
the $K^*\bar{K}$ system that couples to the $\eta(1405)$ and the
$K\bar{K}$ system coupling to the $f_{\, 0}$ can have large on-mass-shell,
isospin-violating contributions.

While our understanding of the low mass scalar mesons remains unclear, 
it seems that detailed studies -- both theoretical and experimental --
of isospin violations in processes involving the $a_{\; 0}(980)$ and $f_0(980)$ 
can provide important probes of their inner workings.  The results
presented above are from data samples that are small fractions of
what we ultimately expect to collect with BESIII.  With the full
data sets we will be able to provide theorists with  precision measurements
of the $a_0(980)\leftrightarrow f_0(980)$ mixing parameters and other quantities
related to these mesons.

\subsection{$\jpsi, \psip \to 3\pi$ and the $\rho\pi$ puzzle}

The oldest puzzle in charmonium physics is the so-called $\rho\pi$ puzzle.
$\jpsi\rt\rho\pi$ is the strongest hadronic decay mode of the $\jpsi$, with
a branching fraction of $(1.69\pm 0.15)$\%~\cite{pdg2010}.  The lowest-order
diagram for this decay is expected to be the three-gluon annihilation process.
The same diagram is expected to
apply to the $\psip$ and, thus, the partial width $\Gamma(\psip\rt\rho\pi)$
is expected to be that for the $\jpsi$, scaled by the ratio of the
$\ccbar$ wavefunctions at the origin and a phase-space factor.  (The ratio
of the wavefunctions at the origin is determined by comparing the
$\jpsi\rt\ee$ and $\psip\rt\ee$ partial widths.)  The result of this reasoning
is the famous ``12\% rule,'' which says that the branching fraction
for $\psip$ to some hadronic state should be (roughly) 12\% that of
the $\jpsi$ to the same final state.  While this simple rule more-or-less
works for many decay modes, it fails miserably for $\psip\rt\rho\pi$
decays, where $Bf(\psip\rt\rho\pi)=(3.2\pm 1.2)\times10^{-5}$, nearly a factor
of a hundred below the 12\%-rule expectation.

\begin{figure}
\mbox{
  \includegraphics[width=0.48\textwidth]{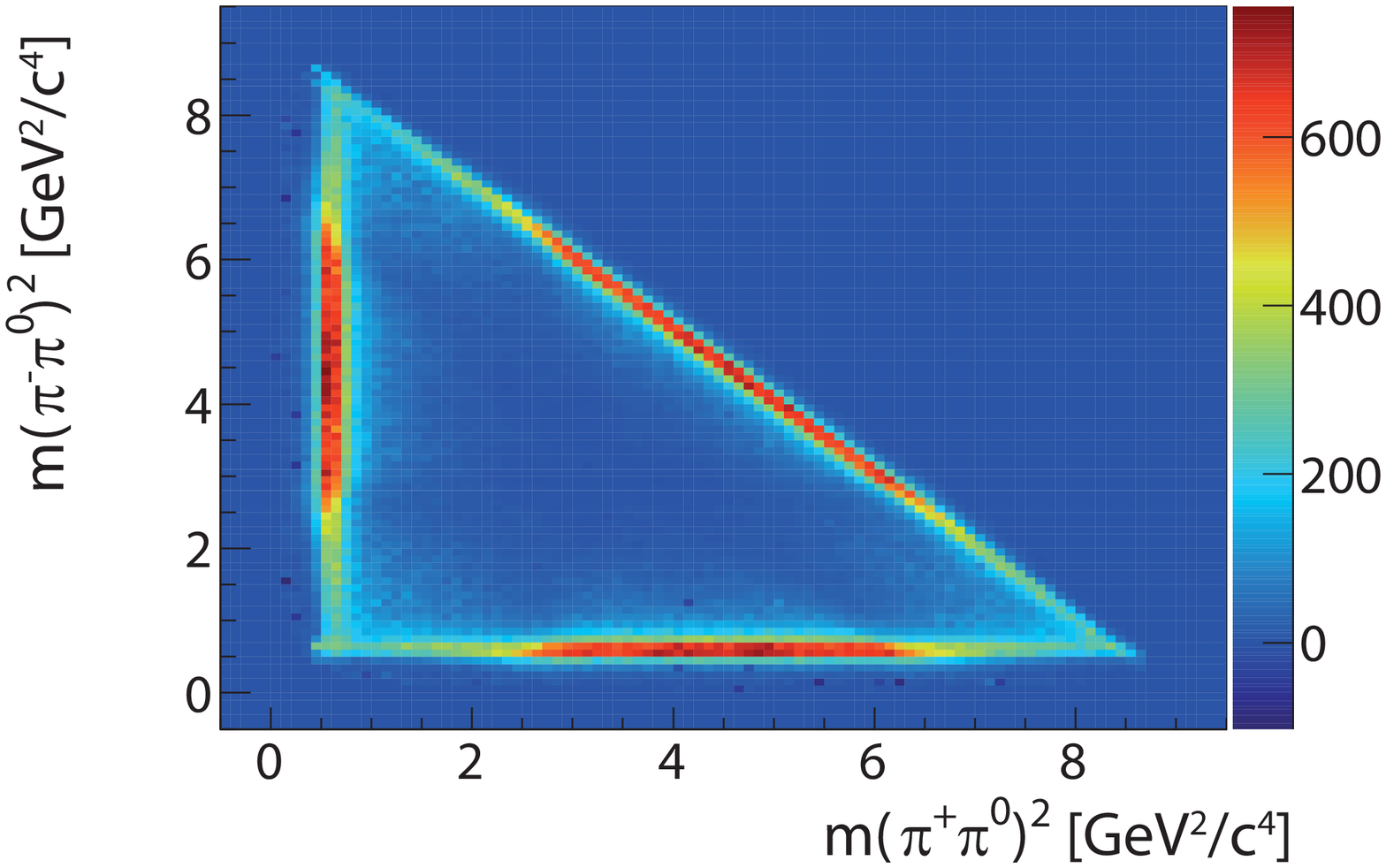}
}
\mbox{
  \includegraphics[width=0.48\textwidth]{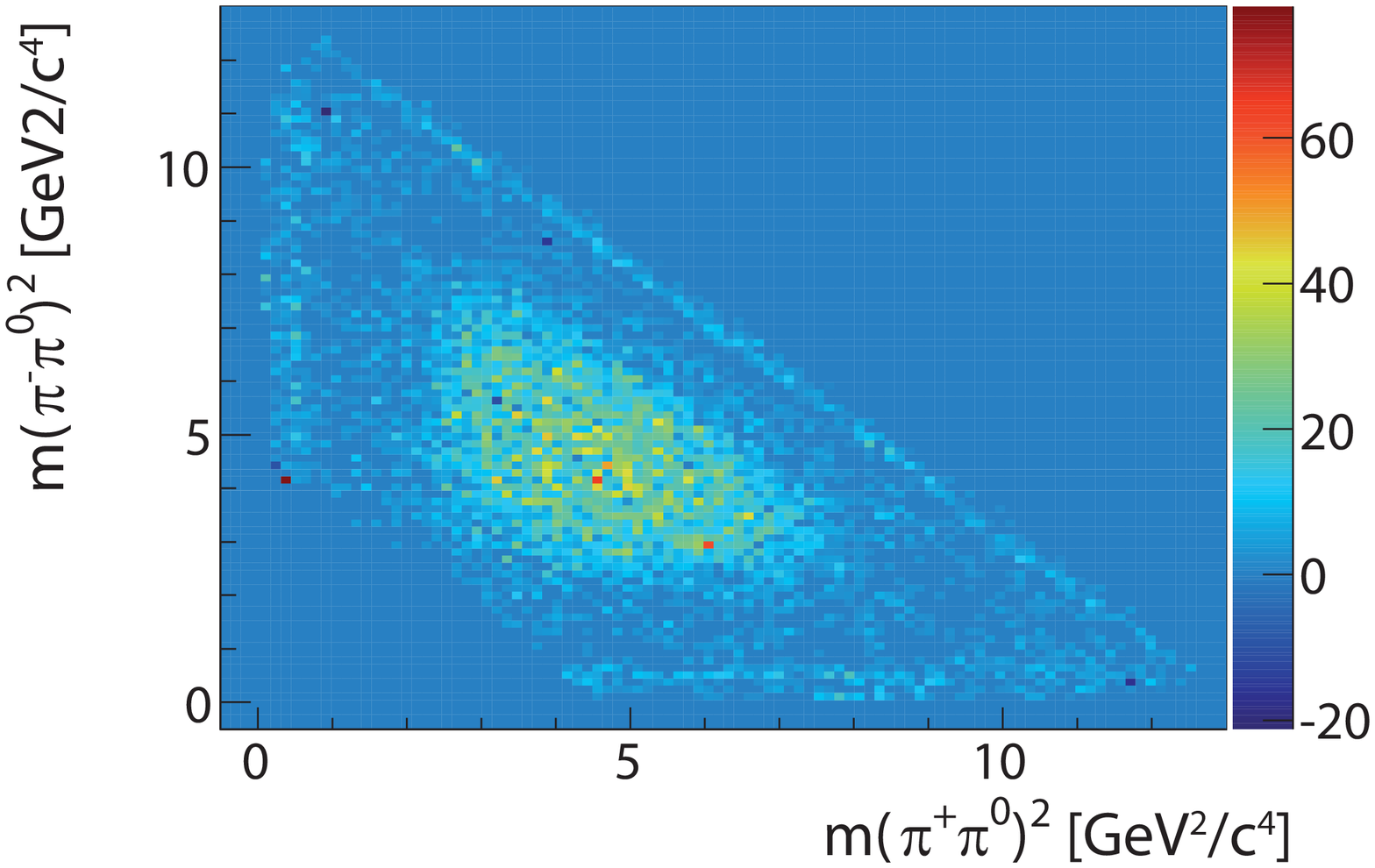}
}
\mbox{
  \includegraphics[width=0.48\textwidth]{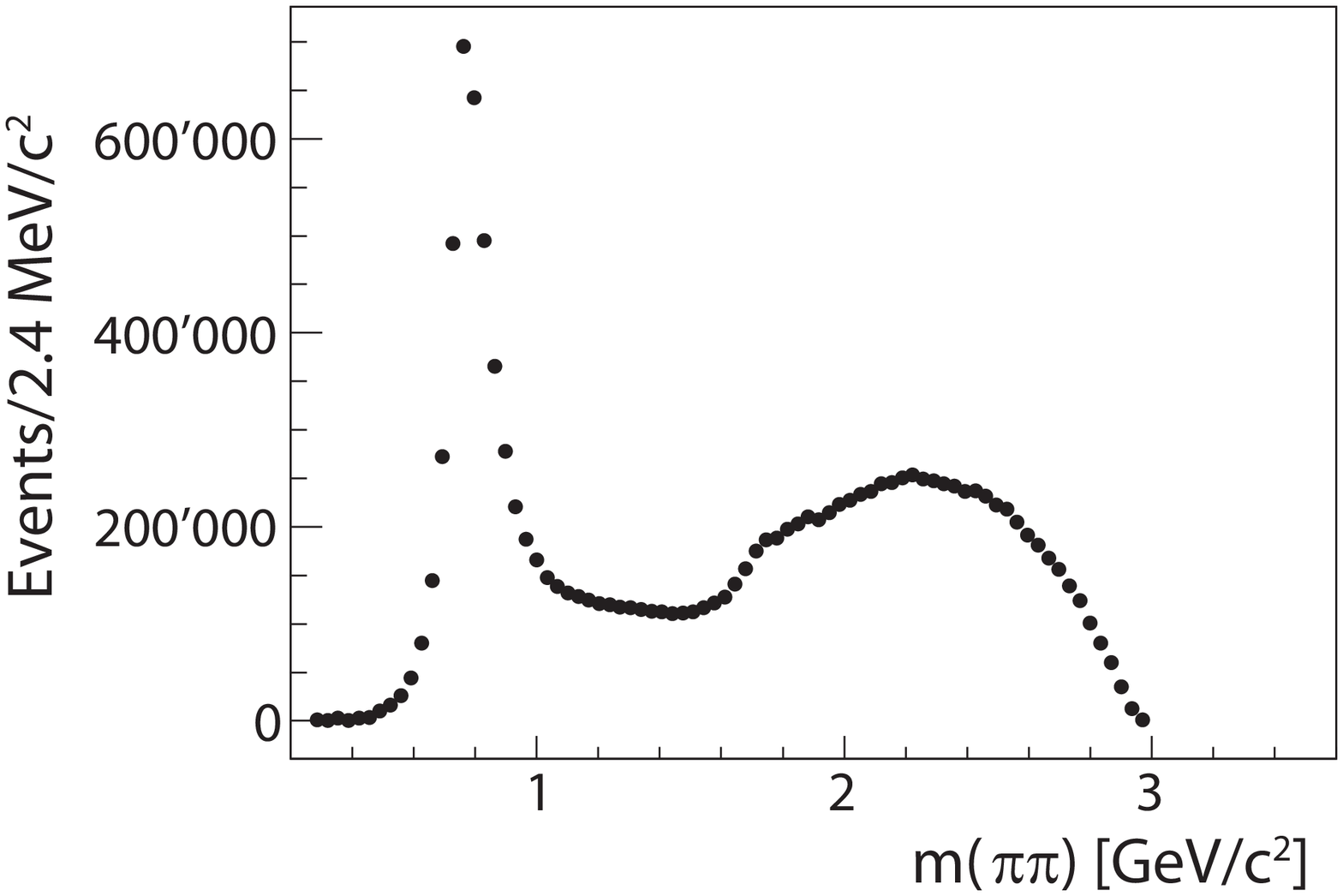}
}
\mbox{
  \includegraphics[width=0.48\textwidth]{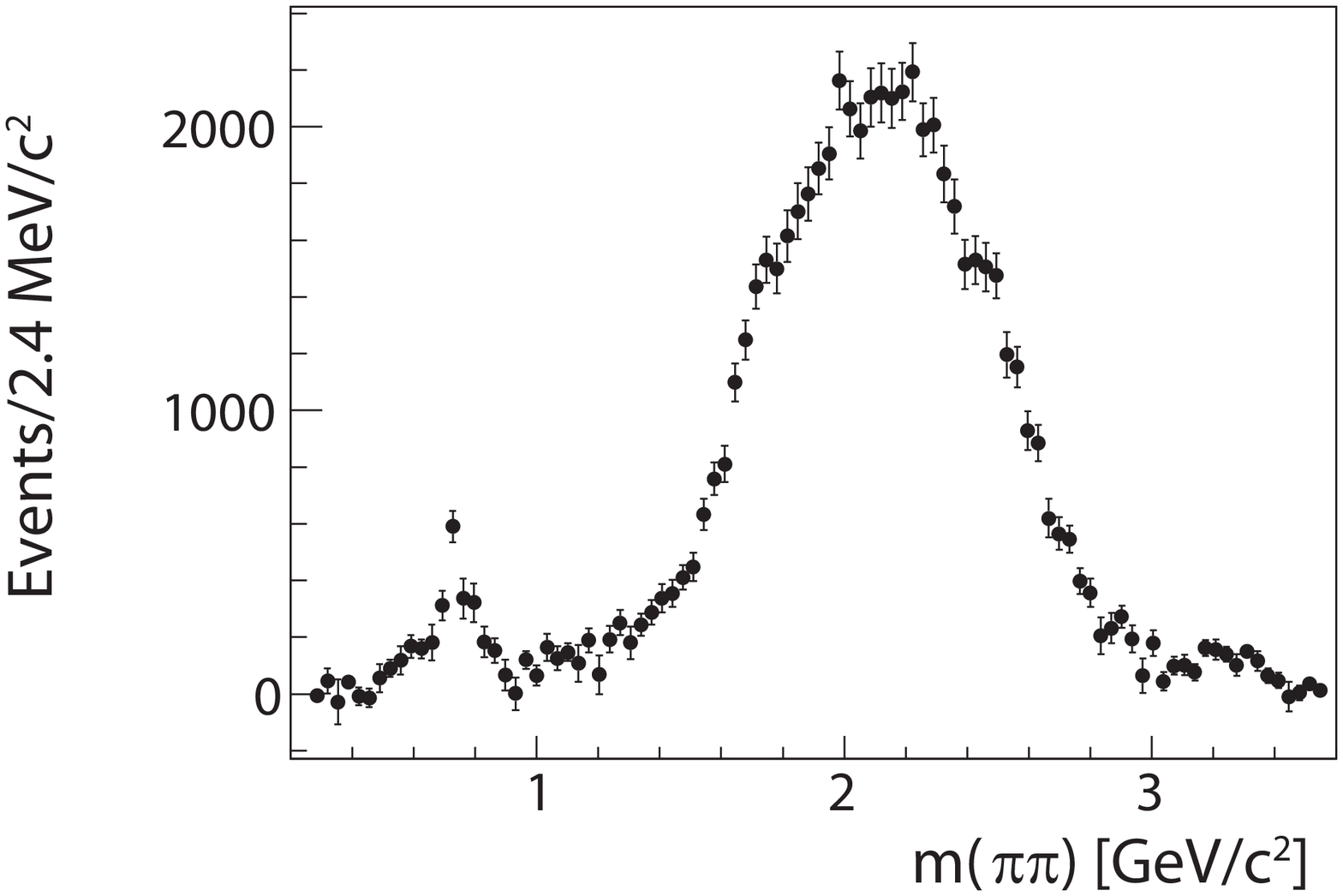}
}
\caption{{\bf Top:} the $M^2(\pi^-\piz)$ (vertical) {\it vs.} $M^2(\pi^+\piz)$
(horizontal) for (left) $\jpsi\rt\pipi\piz$ and 
  (right) $\psip\rt \pipi\piz$ decays.
{\bf Bottom:} the $M(\pi\pi)$ projections of the Dalitz plots. 
}
\label{fig:rhopi}
\end{figure}

BESIII has recently reported on a high-statistics study of of
$\jpsi\rt\pipi\piz$ and $\psip\rt\pipi\piz$~\cite{bes3_rhopi} using the
225M event $\jpsi$ and 106M event $\psip$ data samples. 
The $M^2(\pi^-\piz)$ (vertical) 
{\it vs.} $M^2(\pi^+\piz)$ (horizontal) Dalitz plot
distributions, shown in the top panels of Fig.~\ref{fig:rhopi}, 
for the $\jpsi$ (left) and $\psip$ (right) data samples, could not be more
different.  The center of the $\jpsi\rt\pipi\piz$ Dalitz plot
is completely devoid of events, while in the $\psip\rt\pipi\piz$
plot most of the events are concentrated in the center.  
The dynamics of the two processes are completely different, 
in spite of the fact that the underlying process is expected 
to be very similar.  
The $\rho\pi$ puzzle is becoming even more puzzling.

\subsection{Near-threshold enhancement in $\jpsi\rt\gamma\omega\phi$}

An anomalous enhancement near threshold in the invariant-mass 
spectrum of $\omega\phi$, denoted as $X(1810)$, was first reported 
by the BESII experiment in the decays of $\jpsi\rt\gamma\omega\phi$
with a statistical significance of larger than 10$\sigma$ ~\cite{Ablikim:2006}.
A partial wave analysis (PWA) of BESII data with a helicity 
covariant amplitude showed that the $X(1810)$, with a mass and 
width of $M = 1812^{+19}_{-26}\pm18$ MeV/$c^2$, and
$\Gamma = 105\pm20\pm28$ MeV/$c^2$, respectively, and a product branching
fraction ${\cal B}$($\jpsi\to\gamma$ $X(1810)$) $\cdot$ 
${\cal B}$($X(1810)$$\to\omega\phi$)
 = $[2.61\pm0.27\pm0.65]\times10^{-4}$, 
favors $J^{PC}=0^{++}$ over $J^{PC}=0^{-+}$ or $2^{++}$.
The decay of $\jpsi\rt\gamma\omega\phi$ is a doubly OZI suppressed 
process with a production rate that is suppressed relative to 
$\jpsi\to\gamma\omega\omega$ or $\jpsi\to\gamma\phi\phi$
by at least one-order-of magnitude~\cite{oneorder}. 
The observation of $X(1810)$ has stimulated much theoretical speculation. 
Possible interpretations of $X(1810)$ include a tetraquark state 
(with structure $q^2\overline{q}^2$)~\cite{Bing-An:2006}, 
a hybrid~\cite{Kung-Ta:2006}, a glueball state~\cite{Bicudo:2007}, 
an effect due to the intermediate meson rescattering~\cite{Qiang:2006}, 
a threshold cusp attracting a resonance ~\cite{D.V.:2006}, etc. 
As of now, none of these interpretations has either been 
established or ruled out by experiment. 
A search for the $X(1810)$ was performed by the Belle collaboration 
in the decay of $B^{\pm}\to K^{\pm}\omega\phi$~\cite{belle},
but no obvious $X(1810)$ signal was observed.

\begin{figure}[htbp]
\vskip -0.1 cm \centering
\includegraphics[width=0.48\textwidth]{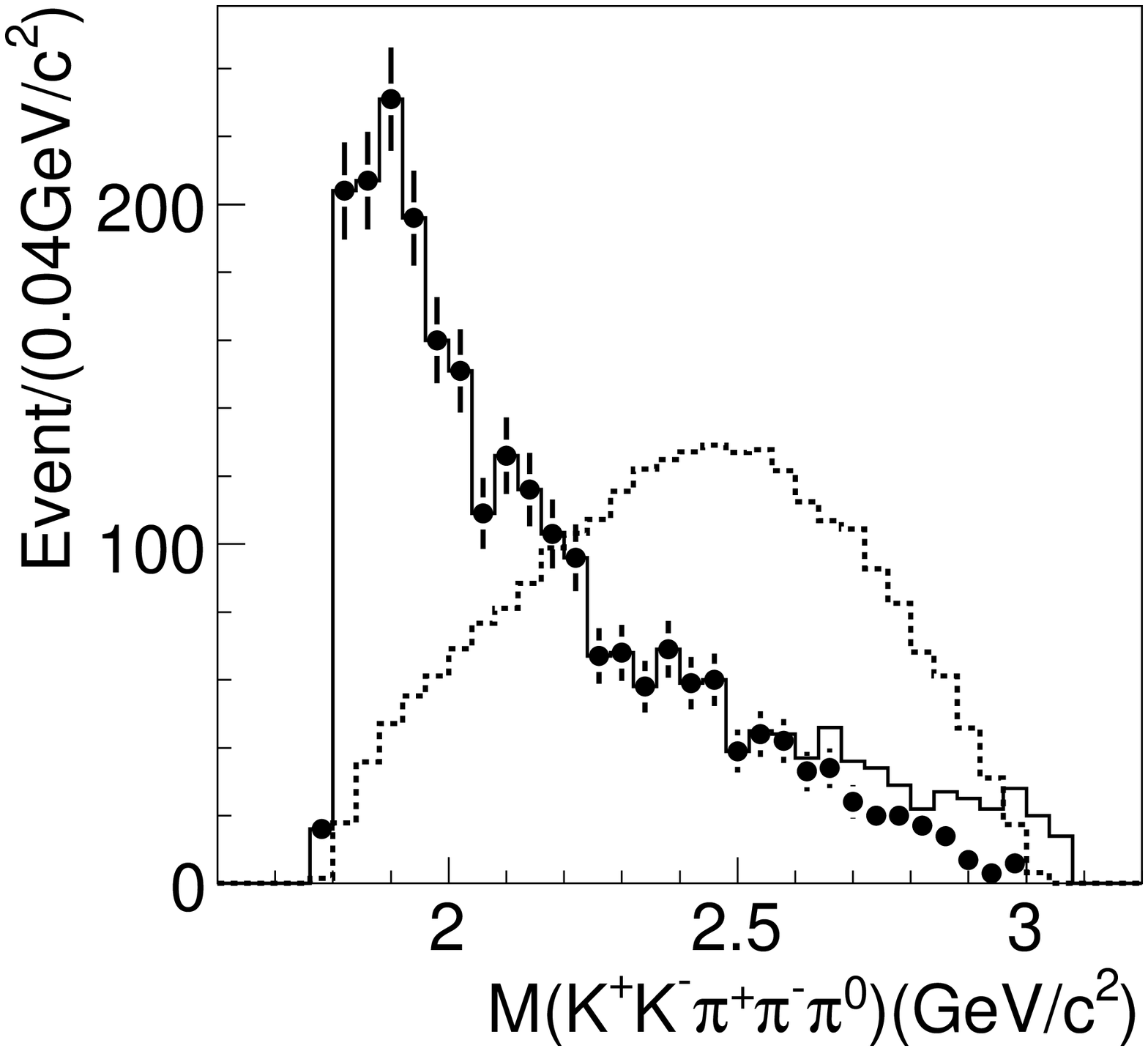}
\put(-25,140){(a)}
\includegraphics[width=0.48\textwidth]{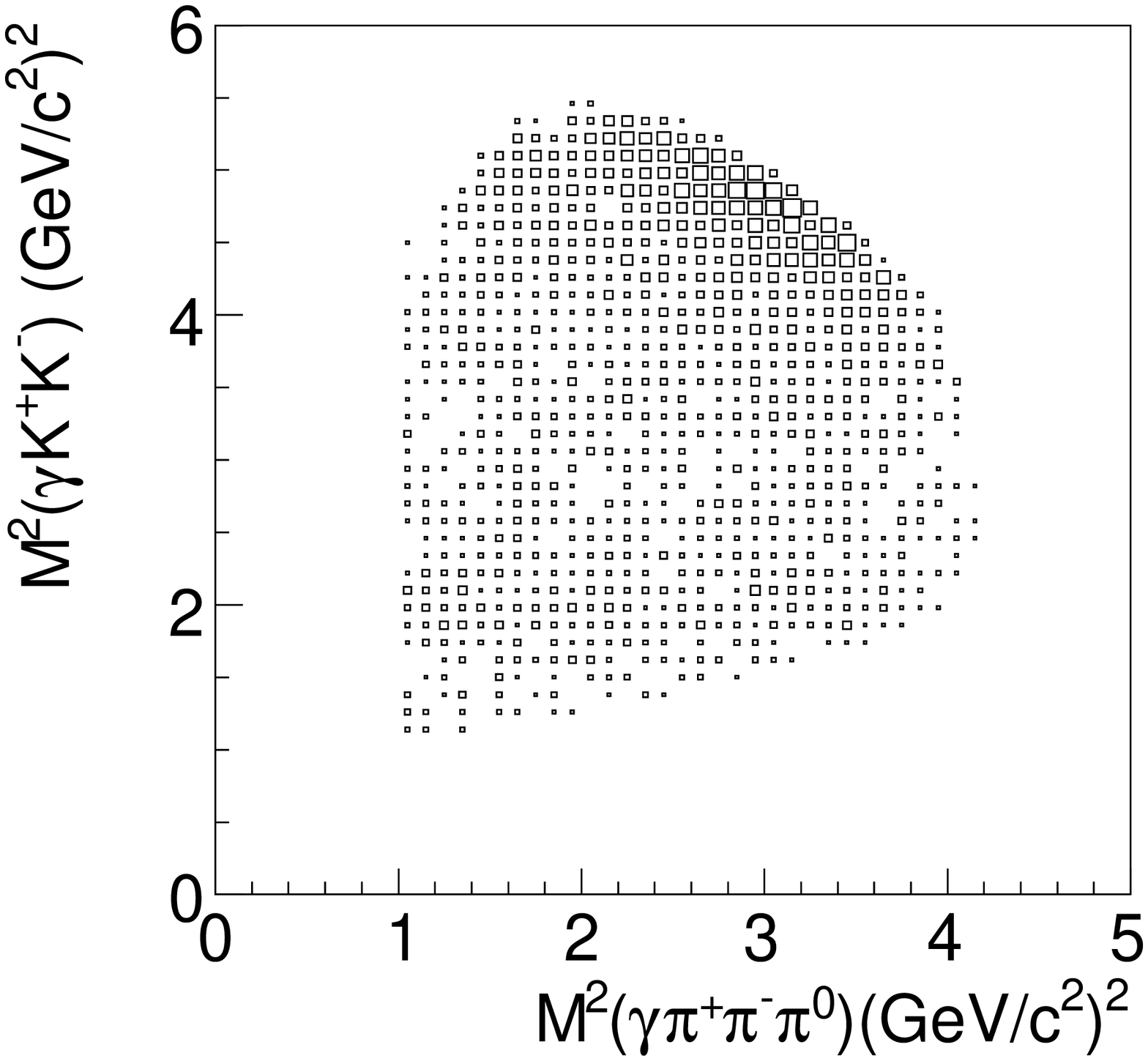}
\put(-25,140){(b)}
\vskip -0.5cm
\caption{(a) The invariant-mass distribution of $\kk\pipi\piz$; 
the dashed line is the mass distribution of the phase space MC sample; 
the solid histogram shows the mass distribution without the requirement 
of $M(\gamma\pipi\piz$)$>$1.0 GeV/$c^{2}$.
(b) Dalitz plot of $M^2(\gamma\pipi\piz)$ {\it versus} $M^2(\gamma\kk)$.}
\label{mwf}
\end{figure}

Using 2.25$\times$10$^8$ $\jpsi$ events, BESIII has re-studied
the decay of $\jpsi\to\gamma\omega\phi$, $\omega\to\pipi\piz$,  
$\phi\to K^+K^-$. The enhancement structure near the $\omega\phi$ 
invariant-mass threshold is confirmed with a statistical significance 
larger than 30$\sigma$. A partial wave analysis with a tensor covariant 
amplitude confirms that the spin-parity of $X(1810)$ is 0$^{++}$.
The mass and width of the $X(1810)$ are determined to be 
$M=1795\pm7^{+23}_{-5}$ MeV/$c^2$ and $\Gamma=95\pm10^{+78}_{-34}$ MeV/$c^2$, 
respectively, and the product branching fraction is
${\cal B}(\jpsi\to\gamma X(1810))\times {\cal B}(X(1810)\to\omega\phi)
=(2.00\pm0.08^{+1.38}_{-1.00})\times10^{-4}$,
where the first error is statistical and the second systematical.
These preliminary results are consistent with those from 
BESII within errors. Fig.~\ref{mwf} shows the invariant mass 
spectrum of $\kk\pipi\piz$ and the Dalitz plot.


\section{Charmonium physics}

\subsection{Meaurements of $h_c$ mass, width and branching fractions}

\begin{figure}
\begin{center}
\includegraphics[width=0.6\textwidth]{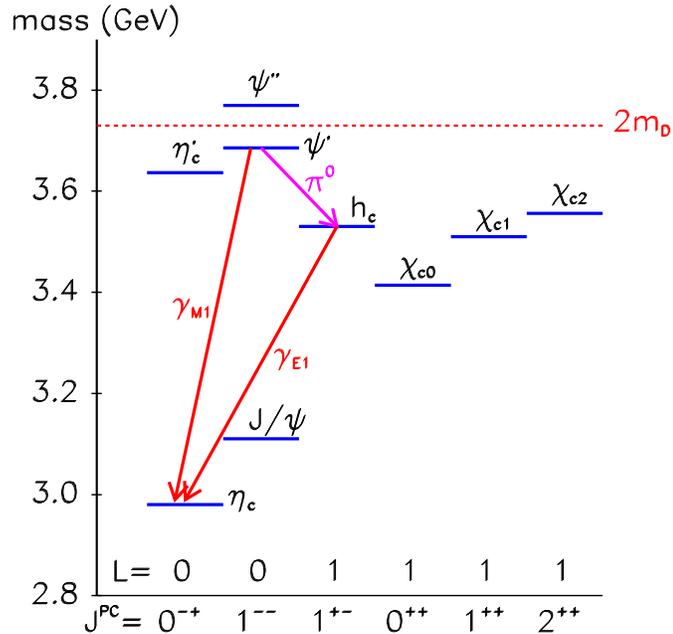}
\caption{The spectrum of the low-lying charmonium mesons.  The red dashed
line indicates the $M=2m_{ D}$ open-charmed threshold. States with mass above
this value can decay to final states containing $D$ and $\bar{D}$ mesons
and are typically broad; states below this threshold are relatively narrow.
The magenta and red arrows indicate transitions used for the $\etac$
and $h_c$ measurements reported here.   
} 
\label{fig:charmonium}
\end{center}
\end{figure}

The charmonium mesons are important because of their simplicity and their accessibility
by a variety of theoretical approaches, including effective field theories and
lattice QCD~\cite{nora}.
Because of their large mass, the charmed quarks bound in the charmonium meson states
have relatively low velocities, $v^2\sim 0.3$, and non-relativistic potential models
can be used with relativisitic effects treated as small perturbations.
With the discovery of the $\etac^{\prime}$ by Belle in 2002~\cite{belle_etacp} and the $h_c$
by CLEO in 2005~\cite{cleo_hc}, all of the charmonium states below the $M=2m_{~D}$ open-charm
threshold have been identified (see Fig.~\ref{fig:charmonium}).  
An experimental task now is the provision of precision
measurements that can challenge the various theories that address this system.

The singlet $P$-wave $h_c$ meson is notoriously difficult to study.  In
fact, despite considerable experimental efforts, it evaded detection
for some thirty years until it was finally seen by CLEO in 2005 in the
isospin-violating $\psip\rt\piz h_c$ transition (indicated by a
magenta arrow in Fig.~\ref{fig:charmonium})~\cite{cleo_hc}.
To date, it has only been seen by two groups, CLEO and
BESIII~\cite{bes3_hc1} and only via the 
strongly suppressed $\psip\rt\piz h_c$ process.

In lowest-order perturbation theory, the $h_c$ mass
is equal to the spin-weighted-average of the
triplet $P$-wave $\chi_{c0,1,2}$ states: 
$<m_{\chi_{cJ}}>=(m_{\chi_{c0}}+3m_{\chi_{c1}}+5m_{\chi_{c2}})/9=3525.30\pm 0.04$~MeV.
Theoretical predictions for the  branching fraction for $\psip\rt\piz h_c$
are in the range $(0.4\sim 1.3)\times 10^{-3}$, the
E1 radiative transition $h_c\rt \gamma\etac$ is expected to
be the dominant decay mode with
a branching fraction somewhere between $40\sim 90$\%, 
and the $h_c$ total width is expected to be less than 1~MeV~\cite{kuang}.

Three detection \& analysis methods have been used to study $h_{\, c}$ production and decay,
all of them use the processes indicated by arrows in 
Fig.~\ref{fig:charmonium}:
\begin{description}

\item[inclusive] In the ``inclusive'' mode, only the $\piz$ is detected
and the $h_{\, c}$ shows up as a peak in the mass recoiling against the
detected $\piz$, which is inferred from conservation of energy and momentum.
The inclusive mode signal yield is proportional the 
$Bf(\psip\rt\piz h_c)$.  This mode has the highest background.

\item[E1-tagged] In the ``E1-tagged'' mode the $\piz$ and the
E1 transition $\gamma$ from the $h_c\rt \gamma \etac$,
with energy in the range $465-535$~MeV, are detected.
The E1-tagged signal yield is proportional to the
branching fraction product 
$Bf(\psip\rt\gamma h_c)\times Bf(h_c\rt\gamma\etac)$.  
The background for this mode is relatively smaller than
that for the inclusive mode.

\item[exclusive] In the ``exclusive'' mode, the $\piz$, E1-$\gamma$
and all of the decay products of the $\etac$ are detected.  Here
all final-state particles are detected and energy-momentum
conserving kinematic fits can be used to improve the resolution.
The backgrounds are small and the yield is proportional
to a triple product of branching fractions, including that
for the $\eta_{\; c}$ decay channel that is detected.

\end{description} 
\noindent
The CLEO observation used both the E1-tagged and exclusive modes.
BESII has reported results from the inclusive and E1-tagged modes;
an exhaustive study of exclusive channels is in progress.

\begin{figure}
\begin{center}
\includegraphics[width=0.6\textwidth]{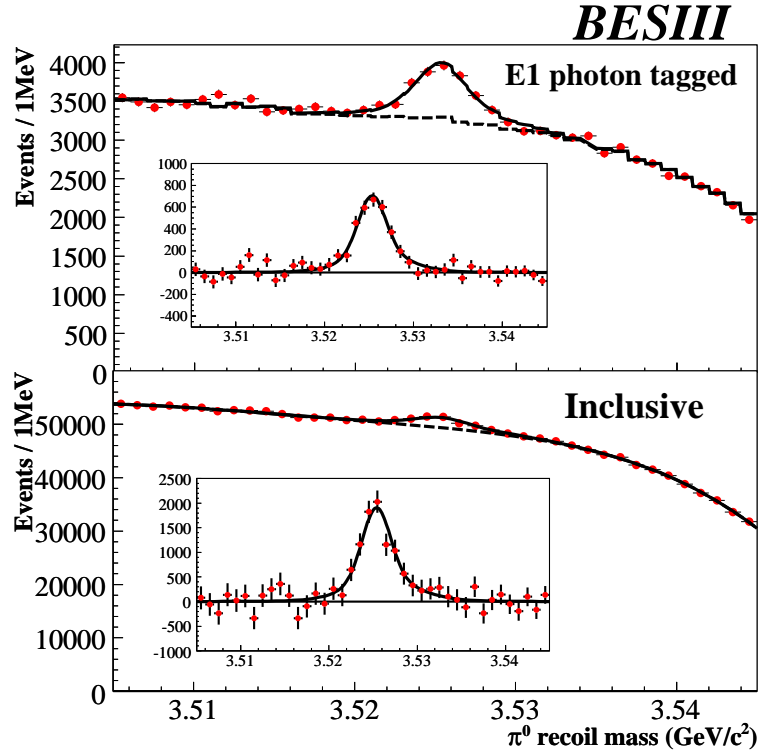}
\caption{The $\piz$ recoil mass for E1-tagged (top) 
and inclusive (bottom)$\psip\rt\piz X$
decays.  The insets show the signal yields with the fitted backgrounds
subtracted. 
} 
\label{fig:hc}
\end{center}
\end{figure}

The BESIII $\piz$ recoil mass distributions for the E1-tagged and
inclusive modes are shown in Fig.~\ref{fig:hc}.  The E1-tagged 
sample (top) has the most distinct signal and this
is used to determine the mass and width of the $h_c$.
The solid curve in the figure is the result of a fit using a BW
function convolved with a MC-determined resolution function to
represent the signal, and a background shape that is determined
from events with no photon in the E1 signal region, but with a
photon in the E1-tag sidebands.  From the fit, the mass and
width are determined to be     
\begin{eqnarray}
m_{h_c} &=& 3525.40 \pm 0.22~{\rm MeV}\\
\Gamma_{h_c} &=& 0.73 \pm 0.53~{\rm MeV};
\end{eqnarray}
the 90\% CL upper limit on the width is $\Gamma_{h_c}<1.44$~MeV.
With this mass value, the $P$-wave hyperfine splitting is 
$<m_{\chi_{cJ}}> - m_{h_c}=-0.10\pm 0.22$~MeV, consistent with zero.
From the signal yield, the product branching fraction
$Bf(\psip\rt\piz h_c)\times Bf(h_c\rt\gamma\etac) = (4.48\pm 0.64)\times 10^{-4}$
is determined.

The inclusive $\piz$ recoil mass distribution is shown in the
lower part of Fig.~\ref{fig:hc}. Here the solid curve is the result
of a fit where the mass and width of the signal function are
fixed at the E1-tagged results and the background is parameterized
by a fourth-order Chebyshev polynomial with all parameters allowed 
to float.  The signal yield and the product branching fraction results
from the E1-tagged mode
are used to make the first determination of the individual branching fractions:
\begin{eqnarray}
Bf(\psip\rt\piz h_c)   &=& (8.4 \pm 1.6)\times 10^{-4}\\
Bf(h_c\rt\gamma\etac ) &=& (54.3 \pm 8.5)\% ,
\end{eqnarray}
which are within the range of theoretical expectations.

\subsection{Meaurement of the $\etac$ mass and width}

The $\etac$ is the ground state of the charmonium system.  The
mass difference between the $\jpsi$ and the $\etac$ is 
due to hyperfine spin-spin interactions and is, therefore,
a quantity of fundamental interest.  However, while the mass
of the $\jpsi$ is known to very high precision --better than
4 PPM-- the $\etac$ mass remains poorly measured, the 2010
PDG world average (WA) value is $m_{\etac}^{~2010}= 2980.3 \pm 1.2$~MeV,
and the measurements that go into this average have poor
internal consistency: the CL of the fit to a single mass
is only 0.0018.  The $\jpsi$-$\etac$ hyperfine mass splitting
derived from this WA is $\delta_{hfs}= 116.6\pm 1.2$~MeV, a
value that has always been above theoretical predictions~\cite{bali}.
The $\etac$ width is also very poorly known; the 2010 PDG WA for
this, $\Gamma_{\etac}^{~2010}=28.6\pm 2.2$~MeV, has a confidence level of
only 0.0001.   

Measurements of the $\etac$ mass and width roughly fall into
two categories, depending on how the $\etac$ mesons used in the
measurement are produced.   Experiments using $\etac$ mesons 
produced via $\jpsi$ radiative transitions tend to find a low
mass ($\sim 2978$~MeV) and narrow width ($\sim 10$~MeV),
while measurements using $\etac$ mesons produced via two-photon
collisions or $B$-meson decays find higher mass and width values.
A primary early goal of the BESIII experiment has been to try
to clear up this situation.

\begin{figure}
\mbox{
  \includegraphics[width=0.48\textwidth]{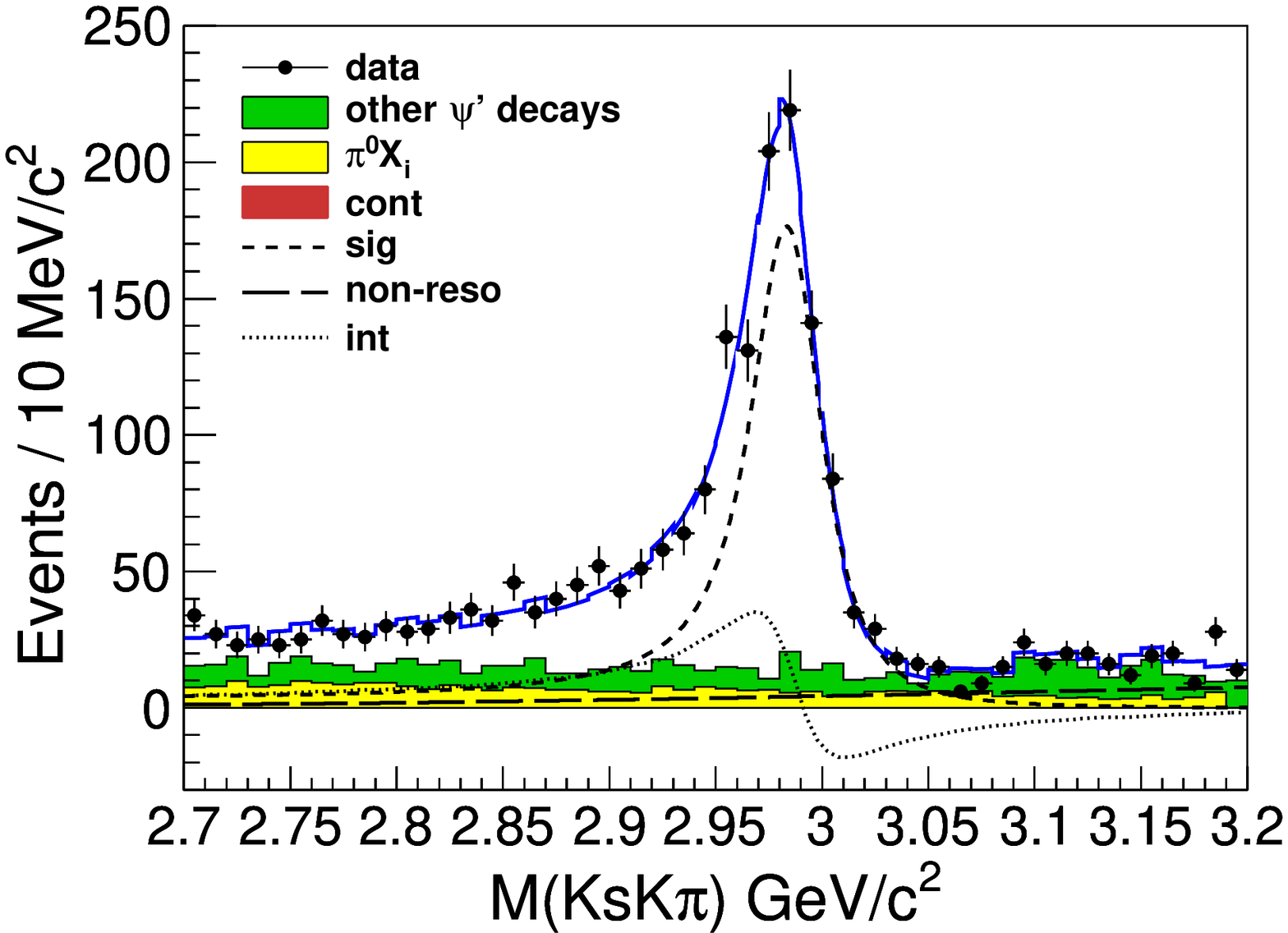}
}
\mbox{
  \includegraphics[width=0.48\textwidth]{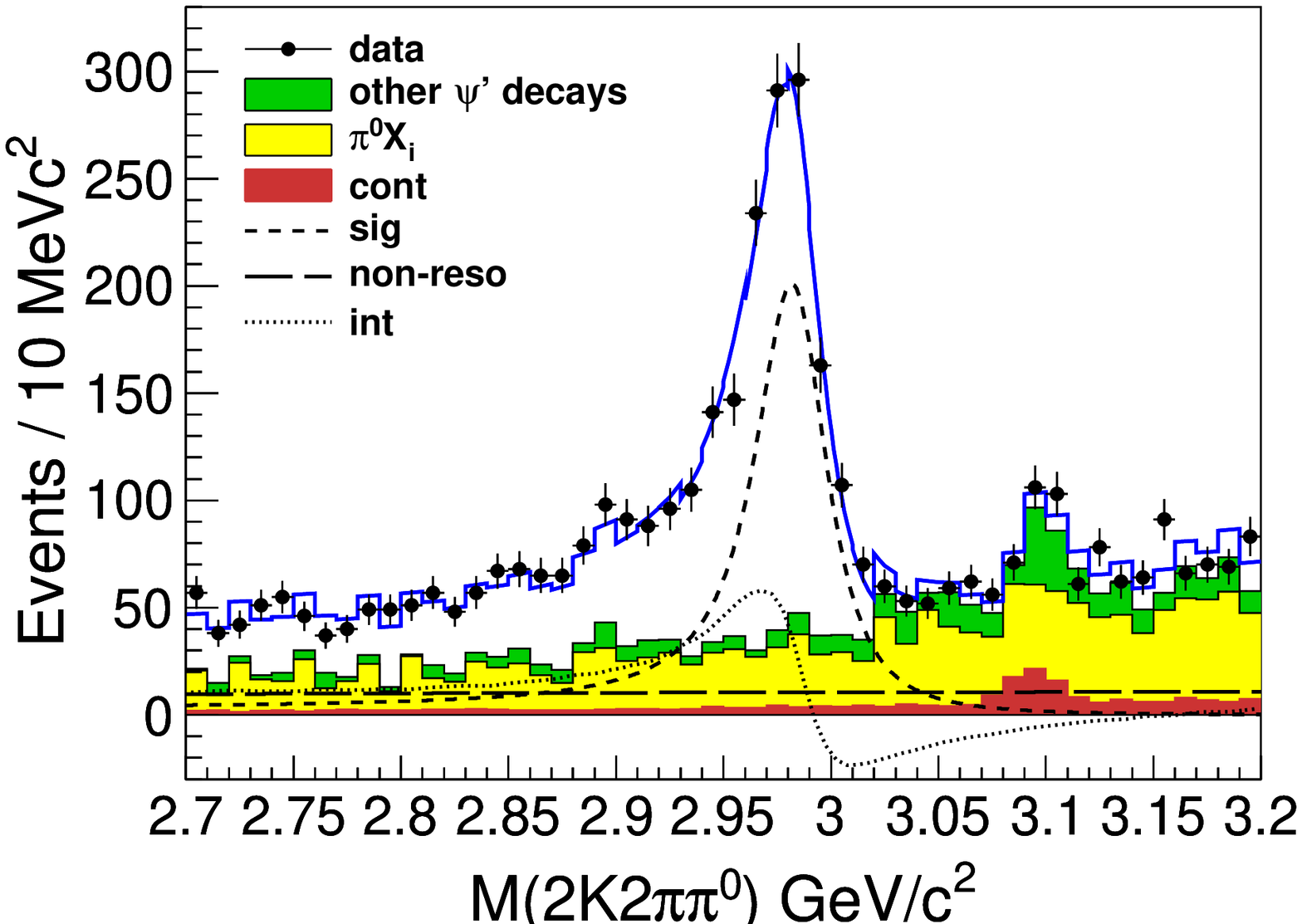}
}
\caption{{\bf Left:} The $K_SK^+\pi^-$ mass spectrum from  
$\psip \rt \gamma K_SK^+\pi-$ decays.
{\bf Right:} The corresponding plot for the $K^+K^-\pipi\piz$ channel.
The main background in most channels, indicated as yellow histograms, are from
$\psip\rt\piz X_i$, where $X_i$ is the same final state 
as the $\etac$ decay mode that is under study, and the $\piz\rt\gamma\gamma$
decay is asymmetric where one $\gamma$ has very low energy and
is not detected. This background is incoherent and does not interfere with
the $\etac$ signal.} 
\label{fig:etac_mass}
\end{figure}

A recently reported BESIII mass and width measurement~\cite{bes3_etac} 
uses samples of $\etac$ mesons produced via the M1 radiative transition
$\psip\rt\gamma\etac$ (indicated by a red arrow in
in Fig.~\ref{fig:charmonium}) that decay to one
of six fully reconstructed final states 
(the inclusion of charge conjugate states is implied):
$\etac\rt X_i$, where $X_i=K_SK^+\pi^-$, $K^+K^-\piz$, $\eta\pipi$,
$K_S K^+\pipi\pi^-$, $K^+K^-\pipi\piz$, and $3(\pipi)$,
where $K_S\rt\pipi$ and $\piz~(\eta)\rt\gamma\gamma$.   Distinct
$\etac$ signals are seen in each of the six channels, two
typical mass spectra are shown in Fig.~\ref{fig:etac_mass}.

In all six channels, the $\etac$ signal has a distinctively asymmetric
shape with a long tail at low masses and a rapid drop on the high mass
side.  This is suggestive of possible interference with a coherent
non-resonant background.   The solid blue curves in Fig.~\ref{fig:charmonium}
show the results of a fit that uses a Breit Wigner (BW) amplitude to
represent the $\etac$ that is weighted by a factor of $E_{\gamma}^7$ that
accounts for the M1 transition matrix element ($E_{\gamma}^3$)
and the wave-function mismatch between the radially excited $\psip$
and the ground-state $\etac$ ($E_{\gamma}^4$); the fit also allows for
interference with background from nonresonant $\psip\rt\gamma X_i$
decays.  Since fits to individual channels give consistent results for
the mass, width and the same value for the interference phase, a
global fit to all six channels at once with a single mass, width
and phase is used to determine the final results:
\begin{eqnarray}
m_{\etac} &=& 2984.3 \pm 0.8~{\rm MeV}\\
\Gamma_{\etac} &=& 32.0 \pm  1.6~{\rm MeV}.
\end{eqnarray}
The value of the phase $\phi$ depends upon whether the constructive or
destructive interference solution is used:  $\phi_{cons}=2.40\pm 0.11$
or $\phi_{des}=4.19\pm 0.09$. (The mass and width values for the
two cases are identical.)  The reason that the interference phase
is the same for all six channels is not understood.

The new BESIII mass and width values agree well with the earlier higher
values found in two-photon and $B$-meson decay meaurements.  The
probable reasons for the low values found by earlier measurements
using $\etac$ mesons produced via radiative charmonium decays are
the effects of the wave-function mismatch~\cite{cleo_etac}
and interference with the
non-resonant background that were not considered.  Using only the new
BESIII $\etac$ mass value, the $\jpsi$-$\etac$ hyperfine mass splitting
becomes smaller:
$\delta_{hfs}=112.6\pm 0.8$~MeV, and in better agreement with theory. 

\subsection{First observation of the M1 transition $\psip\rt\gamma\eta_c(2S)$}

The $\etacp$ was first observed by the Belle collaboration in the
process $B^\pm\to K^\pm \etacp$, $\etacp\to K_S^0K^\pm
\pi^\mp$~\cite{babar_B2Ketacp}.  It was confirmed in the
two-photon production of $\KsKpi$~\cite{babar_2gamEtacp,
cleo_2gamEtacp}, and in the double-charmonium production process
$e^+e^-\to J/\psi c\bar{c}$~\cite{belle_Jpsiccbar, babar_Jpsiccbar}.
The production of the $\etacp$ through a radiative transition from 
the $\psip$ requires a charmed-quark spin-flip and,
thus, proceeds via a magnetic dipole (M1) transition.

\begin{figure*}[htb]
\centering
  \includegraphics[width=0.49\textwidth]{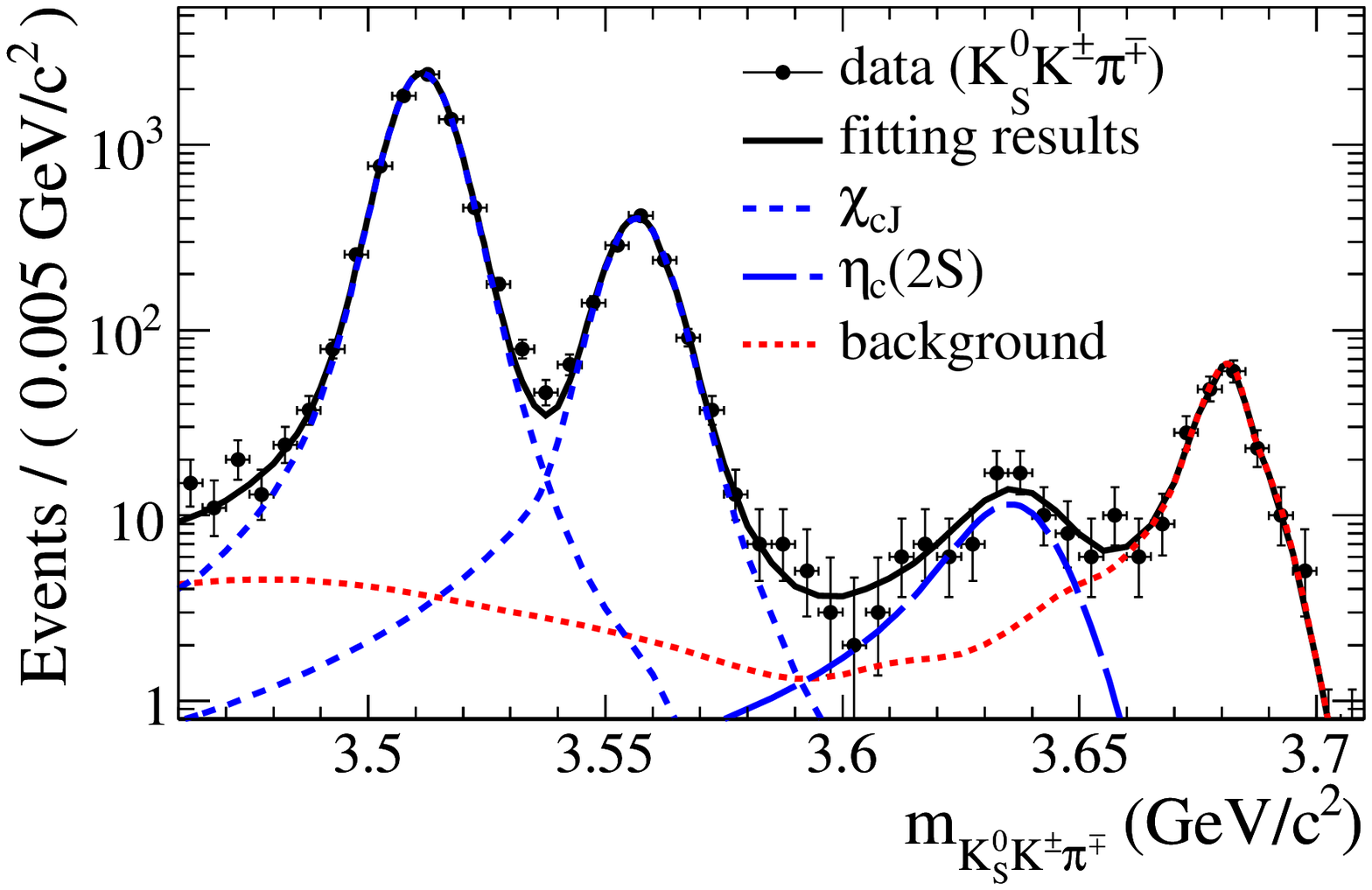}
  \includegraphics[width=0.49\textwidth]{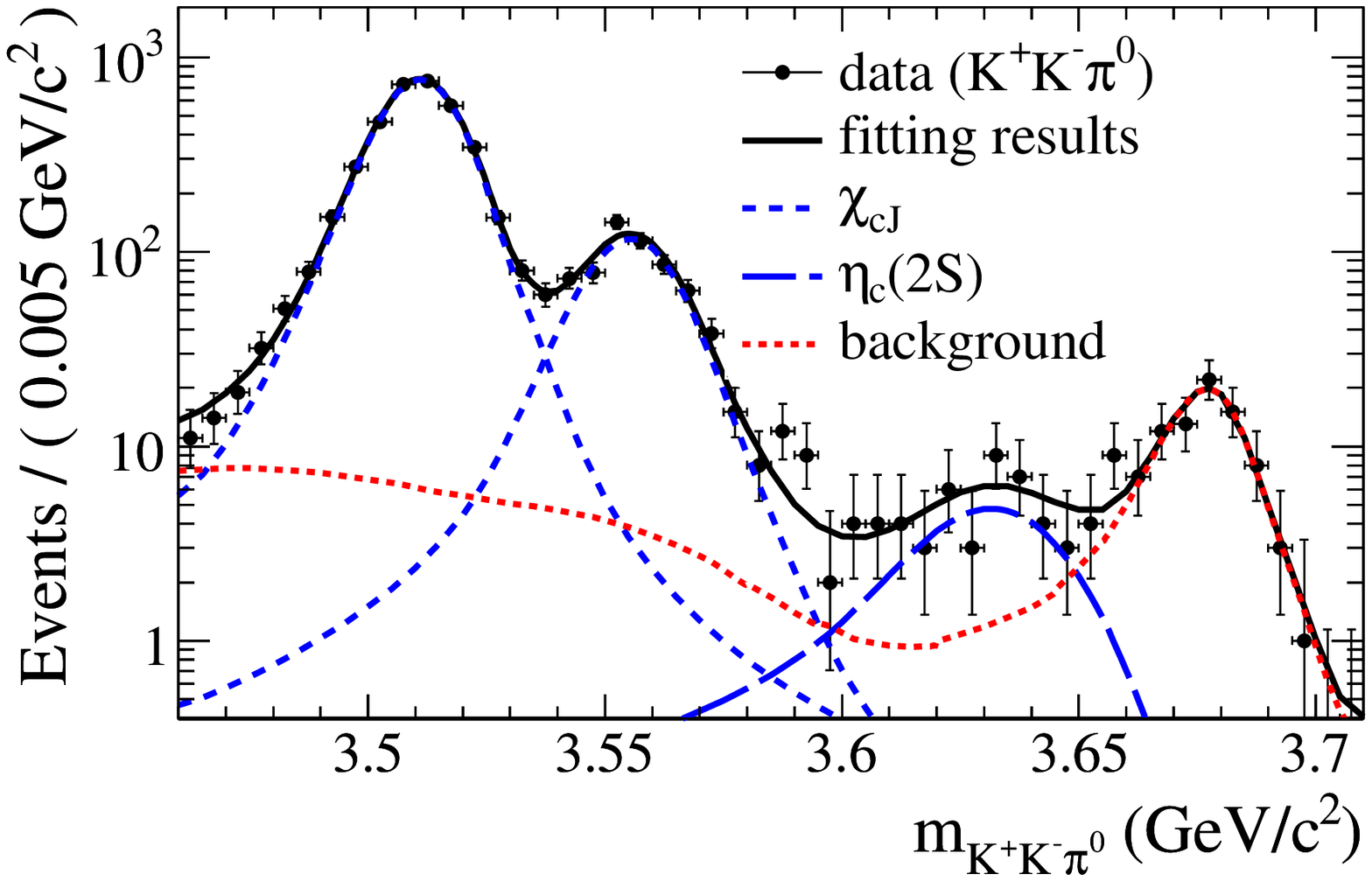}
\caption{The invariant-mass spectrum for $\KsKpi$ (left panel),
$\kkpiz$ (right panel), and the simultaneous likelihood fit to the
three resonances and combined background sources as described in
the text.} \label{pic_fit_etacp}
\end{figure*}

BESIII has reported the first observation of $\psip\to \gamma\etacp$, 
with $\etacp\to \KsKpi$ and $K^+K^-\pi^0$ using 106 M $\psip$ events
~\cite{bes3etacp}.
The mass spectra for the $\KsKpi$ and $\kkpiz$ channels and 
a simultaneous fit to extract the yield, mass and width of $\etacp$
are shown in Fig.~\ref{pic_fit_etacp}.

Analyses of the processes $\psip\to
\gamma\etacp$ with $\etacp\to \KsKpi$ and $\kkpiz$ gave an
$\etacp$ signal with a statistical significance of greater than 10
standard deviations under a wide range of assumptions about the
signal and background properties.  The data are used to obtain
measurements of the $\etacp$ mass ($M(\etacp)=3637.6\pm
2.9_\mathrm{stat}\pm 1.6_\mathrm{sys}$~MeV/$c^2$), width
($\Gamma(\etacp)=16.9\pm 6.4_\mathrm{stat}\pm
4.8_\mathrm{sys}$~MeV), and the product branching fraction 
($\BR(\psip\to \gamma\etacp)\times \BR(\etacp\to
K\bar K\pi) = (1.30\pm 0.20_\mathrm{stat}\pm
0.30_\mathrm{sys})\times 10^{-5}$). Combining our result with a
BaBar measurement of $\BR(\etacp\to K\bar K \pi)$, we find the
branching fraction of the M1 transition to be
$\BR(\psip\to\gamma\etacp) = (6.8\pm 1.1_\mathrm{stat}\pm
4.5_\mathrm{sys})\times 10^{-4}$.

\subsection{Evidence for $\psip\rt\gamma\gamma\jpsi$}

Two-photon spectroscopy has been a very powerful tool for the study
of the excitation spectra of a variety of systems with a wide range of sizes,
such as molecules, atomic hydrogen and positronium~\cite{Pachucki:1996jw}.
Studying the analagous process in quarkonium states is a natural extension
of this work, in order to gain insight into non-perturbative QCD phenomena.
But so far, two-photon transitions in quarkonia have
eluded experimental observation~\cite{Bai:2004cg,Adam:2005uh,:2008kb}.
For example, in a study of $\psip\to\gm\chi_{cJ}(J=0,1,2)$ reported by
CLEO-c~\cite{:2008kb}, the upper limit for $\br{\psip\to\gm\gm\jpsi}$ was
estimated to be $1\times10^{-3}$.

\begin{figure}[tp!]
\centering
\includegraphics[width=0.5\linewidth]{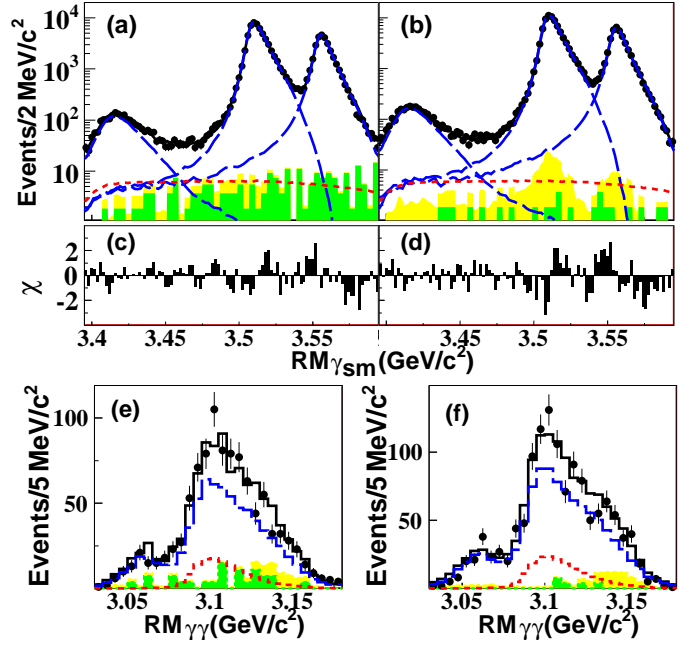}
\caption{(color online) Plot a(b): unbinned maximum likelihood fit to the
   distribution of $RM_{\gmsm}$ in data for $\gm\gm\ee$($\gm\gm\mm$) mode.
   Thick lines are the sum of the fitting models and long-dashed
   lines the $\chicj$ shapes. Short-dashed lines represent
   the two-photon signal processes. Shaded histograms are $\psip$-decay backgrounds (yellow) and non-$\psip$ backgrounds (green), with the fixed amplitude and shape taken from MC simulation and continuum data. Plot c(d): the number of standard deviations, $\chi$, of data points from the fitted curves in plot a(b).
   The rates of the signal process and sequential $\chicj$ processes are derived from these fits.
   Plot e(f): distributions of $RM_{\gm\gm}$ in data, the signal process and known backgrounds with kinematic requirement $3.44\gevcc<RM_{\gmsm}<3.48\gevcc$ and with removal of $RM_{\gm\gm}$ window requirement for $\gm\gm\ee$($\gm\gm\mm$) mode. }
\vspace{-0.5cm}
\label{fig:global_fit}
\end{figure}

BESIII has reported the first evidence for the two-photon transition
$\psip\to\gm\gm\jpsi$, studies of the orientation of the $\psip$ decay plane
and the $\jpsi$ polarization in the decay, as well as the branching
fractions of double $E1$ transitions $\psip\to\gm(\gm\jpsi)_{\chicj}$
through $\chicj$ intermediate states.  The yield of the signal process
$\psip\to\gm\gm\jpsi$, together with those of the cascade $E1$ transition
processes, is estimated by a global fit to the spectrum of $RM_{\gmsm}$. 
The fit results are shown in Fig.~\ref{fig:global_fit}.  The
$\psip\to\gm\gm\jpsi$ transition is observed with a statistical significance
of 6.6$\sigma$. When the systematic uncertainties are taken into account 
with the assumption of Gaussian distributions, the significance is 
evaluated to be 3.8$\sigma$.
A cross-check on the procedures is performed with the $RM_{\gm\gm}$ spectrum
for the events in the region $3.44\gevcc<RM_{\gmsm}<3.48\gevcc$ without
restriction on $RM_{\gm\gm}$, as shown in Fig.~\ref{fig:global_fit}(e) and
(f).  An excess of data above known backgrounds can be seen around the
$\jpsi$ nominal mass, which is expected from the sought-after two-photon
process.  With the inclusion of the estimated yields of the signal process,
the excess is well understood.
The branching fraction of $\psip\to\gm\gm\jpsi$ is determined to be 
$(3.3\pm0.6(\unit{stat})^{+0.8}_{-1.1}(\unit{syst}))\times10^{-4}$ 
(preliminary) using $\jpsi\to\ee$ and $\jpsi\to\mm$ 
decays~\cite{bes3_psip2ggjpsi}.

\section{Charmed meson physics}

The primary goal of the BESIII program is precision studies of weak decay
processes of $D$ and $D_s$ mesons.  The initial phase of this program was
a long data-taking run that accumulated 2.9~fb$^{-1}$ 
at the peak of the $\psi(3770)$ charmonium meson.
This is a resonance in the $\ee\rt \DDbar$ channel with a peak cross section
of about 6~nb at a c.m. energy that is about 40~MeV
above the $E_{c.m.}= 2m_{\, D}$ open-charm mass threshold.  
The $\psi(3770)$ is included in the sketch of the charmonium spectrum 
shown in Fig.~\ref{fig:charmonium}.

The 2.9~fb$^{-1}$ data sample that has already been collected
contains almost 20M $\DDbar$ meson pairs and
is about three times the world's previous largest $\psi(3770)$
event sample collected by CLEO-c and is currently being used for numerous
analyses aimed at searches for rare decays and new physics, 
and improving on the precision of previous measurements.  
In many of the latter cases, the measurements are of
form-factors that are accessible in lattice QCD calculations.  
As the precision of lattice QCD improves, BESIII will provide
more precise measurements that continue to challenge the theory.  

\subsection{$D$ tagging technique}
\begin{figure}[htbp]
\centerline{
\includegraphics[width=12.0cm,height=8.0cm]{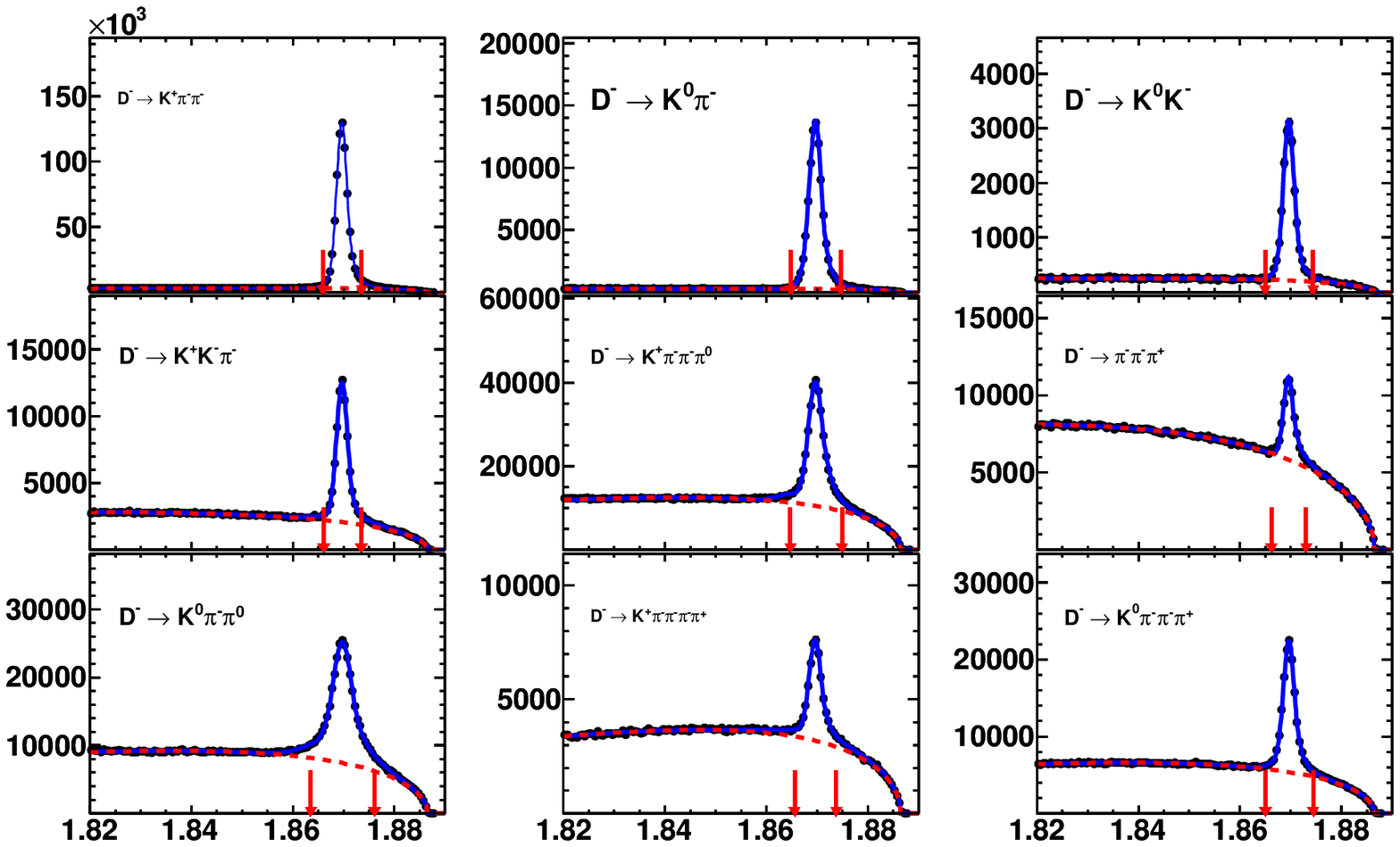}
\put(-212.0,5.0){{M$_{\rm B}$ ~~[GeV/$c^2$]}}
\put(-350.0,80){\rotatebox{90}{Number of Events}}
\put(-300.0,160.0){(a)}
\put(-195.0,160.0){(b)}
\put(-95.0, 160.0){(c)}
\put(-300.0,100.0){(d)}
\put(-195.0,100.0){(e)}
\put(-95.0, 100.0){(f)}
\put(-300.0,50.0){(g)}
\put(-195.0,50.0){(h)}
\put(-95.0, 50.0){(i)}
        }
\caption{Distributions of the beam energy constraint masses
of the $mKn\pi$
combinations for the 9 single tag modes from the data;
where (a), (b), (c), (d), (e), (f), (g), (h), (i)
are for the modes of $D^- \rightarrow K^+\pi^-\pi^-$,
$D^- \rightarrow K^0_s\pi^-$,
$D^- \rightarrow K^0_s K^-$,
$D^- \rightarrow K^+K^-\pi^-$,
$D^- \rightarrow K^+\pi^-\pi^-\pi^0$,
$D^- \rightarrow \pi^+\pi^-\pi^-$,
$D^- \rightarrow K^0_s\pi^-\pi^0$,
$D^- \rightarrow K^+\pi^-\pi^-\pi^-\pi^+$, and
$D^- \rightarrow K^0_s\pi^-\pi^-\pi^+$,
respectively.
}
\label{fig:mbc}
\end{figure}

Most of $\psi(3770)$ decays are to $\DDbar$ meson pairs and nothing else,
because there is not enough enough c.m. energy to produce any other 
accompanying hadrons.  
As a result, The energy of each $D$ meson is half of the total c.m. energy, 
which is precisely known.  Thus, when a $D$ meson is reconstructed in an event,
the recoil system is ``tagged'' as a~$\bar{D}$, and the
constraint on the energy results in reconstructed $D$-meson mass signals
that have excellent resolution ($\sigma= 1.3$~MeV for all charged modes
and $\sigma=1.9$~MeV for modes with one $\piz$) and signal to noise.  
As examples, $D^-$ meson signals for nine commonly used tag decay modes 
are shown in Fig.~\ref{fig:mbc}. 
A maximum likelihood fit to the mass spectrum with a Crystal Ball 
function plus an Gaussian function for the $D^-$ signal and
the ARGUS function to describe background yields the number of 
the singly tagged $D^-$ events for each of the nine modes.
Selecting these candidates for $D^-$ tags within the range marked 
by arrows in Fig.~\ref{fig:mbc} reduce signal number by about $2\%$ 
giving a total of $1586056 \pm 2327$ $D^-$ tags.
In these $D^-$ tags, 20103 $D^-$ tags are reconstructed in more than 
one single $D^-$ tag mode. Subtracting this number of the double 
counting $D^-$ tags from the $1586056 \pm 2327$ $D^-$ tags
yields $1565953 \pm 2327$ $D^-$ tags which are used for further 
analysis of measuring the branching fraction for 
$D^+ \rightarrow \mu^+\nu_{\mu}$ decays.
Moreover, the $\DDbar$ system is in a coherent, $P$-wave
quantum state with $J^{PC}=1^{--}$.  This coherence is unique to $D$ mesons
originating from $\psi(3770)\rt \DDbar$ decays and
permits a number of interesting measurements~\cite{asner}.  For example,
if one $D$ meson is tagged in a pure $CP$ decay mode (such as $K^+K^-$, $\pipi$
or $K_S\piz\piz$ for $CP=+1$, and  $K_S\piz$, $K_S\eta$ or $K_S\omega$ for
$CP=-1$), the decay of the accompanying $D$ meson to a $CP$ eigenstate with the same
$CP$ eigenvalue would be an unambiguous signal for $CP$ violation.

\subsection{Leptonic decays of $D^+$ meson}

In the SM (Standard Model) of particle physics, the $D^+$
meson can decay into $l^+\nu_l$ (where $l$ is $e$, $\mu$ or $\tau$)
through a virtual $W^+$ boson.
The decay rate is determined by the wavefunction overlap of
the two quarks at the origin, and is parameterized by the 
$D^+$ decay constant, $f_{D^+}$.
To the lowest order, as the analogue of the decay width of 
$\pi^+ \rightarrow l^+\nu_l$, the decay width of 
$D^+ \rightarrow l^+\nu_l$ is given by
\begin{equation}
\Gamma(D^+ \rightarrow l^+\nu_{l})=
     \frac{G^2_F f^2_{D^+}} {8\pi}
      \mid V_{\rm cd} \mid^2
      m^2_l m_{D^+}
    \left (1- \frac{m^2_l}{m^2_{D^+}}\right )^2,
\label{eq01}
\end{equation}
where $G_F$ is the Fermi coupling constant, $V_{\rm cd}$ is the
Cabibbo-Kobayashi-Maskawa (CKM) matrix element between the two 
quarks $c\bar d$~\cite{pdg2010} in $D^+$, 
$m_l$ is the mass of the lepton, and
$m_{D^+}$ is the $D^+$ mass.
By measuring the branching fraction of $D^+ \rightarrow \mu^+\nu_{\mu}$,
the decay constant $f_{D^+}$ can be determined.

Candidate events for the decay $D^+ \rightarrow \mu^+\nu_{\mu}$
are selected from the surviving charged tracks in the system 
recoiling against the singly tagged $D^-$ mesons.
To select the $D^+ \rightarrow \mu^+ \nu_{\mu}$,
it is required that there be a single charged track
originating from the interaction region in the system recoiling 
against the $D^-$ tag and the charged track satisfies 
$|\rm {cos \theta}|<0.93$ as well as it is identified as a $\mu^+$.
The $\mu^+$ can be well identified with the passage length
of a charged particle through the MUC since a charged hadron (pion or kaon)
quickly loses its energy due to its strong interactions with the 
absorber of the MUC and most of the hadrons stop in the absorber 
before passing a long passage length in the MUC.
For the candidate event, no extra good photon which is not used 
in the reconstruction of the singly tagged $D^-$ meson
is allowed to be present in the event, where the ``good photon'' 
is the one with deposited energy in the EMC being greater than 300 MeV.

Since there is a missing neutrino
in the purely leptonic decay event, the event should
be characteristic with missing energy $E_{miss}$ and
missing momentum $p_{miss}$
which are carried away by the neutrino.
So they infer the existence of the neutrino by requiring a measured
value of the missing mass squared $M^2_{\rm miss}$
to be around zero. The missing mass squared $M^2_{\rm miss}$
is defined as
\begin{equation}
M^2_{\rm miss} = (E_{\rm beam}-E_{\mu^+})^2 - (- \vec p_{D^-_{\rm tag}}- \vec p_{\mu^+} )^2,
\label{eq_miss2}
\end{equation}
where $E_{\mu^+}$ and $\vec p_{\mu^+}$ are, respectively, the energy and three-momentum of the $\mu^+$,
and $\vec p_{D^-_{\rm tag}}$ is three-momentum of the candidate for $D^-$ tag.

\begin{figure}[htbp]
\centerline{
\includegraphics[width=12.0cm,height=11.0cm]{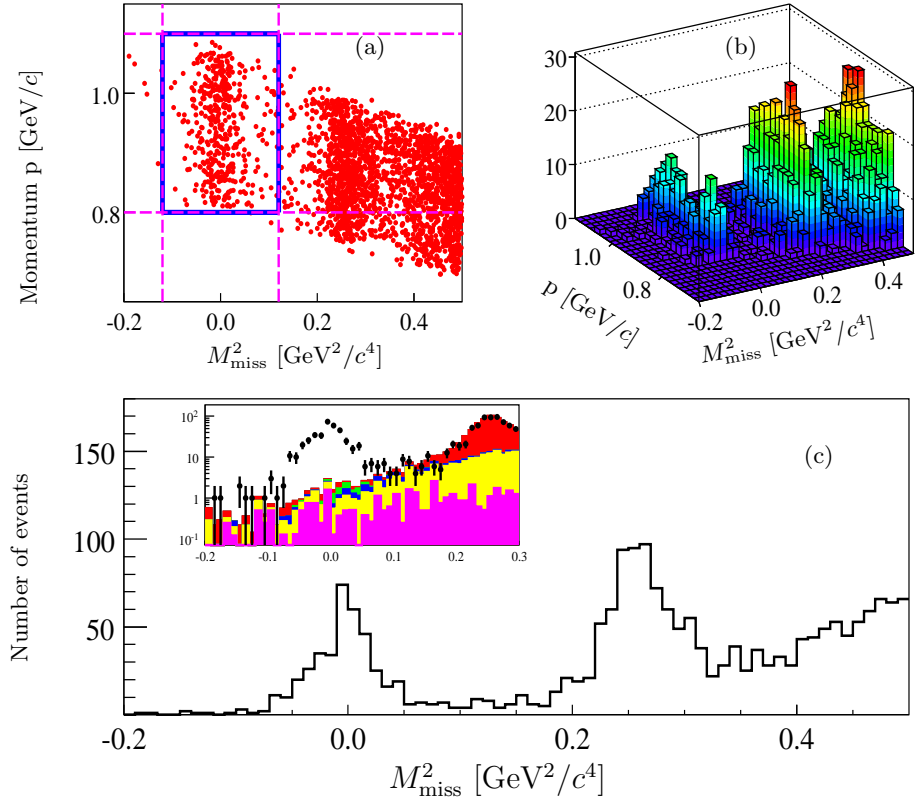}
\put(-276.0,158.0){$M^2_{\rm miss}$~[GeV$^2$/$c^4$]}
\put(-347.0,180){\rotatebox{90}{Momentum p [GeV/$c$]}}
\put(-90.0,158.0){\rotatebox{10}{\small $M^2_{\rm miss}$~[GeV$^2$/$c^4$]}}
\put(-150.0,186){\rotatebox{-32}{\small p [GeV/$c$] }}
\put(-207.0,-3.0){\large{$M^2_{\rm miss}$~[GeV$^2$/$c^4$]}}
\put(-350.0,45){\rotatebox{90}{Number of events}}
\put(-220.0,275.0){(a)}
\put(-80.0,275.0){(b)}
\put(-50.0, 120.0){(c)}
           }
\caption{Distributions of $M^2_{\rm miss}$, where (a) and (b) are scatter plots
of the identified muon momentum $p$ VS $M^2_{\rm miss}$,
and (c) is the distribution of $M^2_{\rm miss}$.
The insert shows the signal region for $D^+ \rightarrow \mu^+\nu_{\mu}$
on a log scale, where dots with error bars are for the data,
histograms are for the simulated backgrounds from
$D^+ \rightarrow K^0_L \pi^+$ (red),
$D^+ \rightarrow \pi^0\pi^+$ (green),
$D^+ \rightarrow \tau^+ \nu_{\tau}$ (blue) and
other decays of $D$ mesons (yellow) as well as
from $e^+e^- \rightarrow$non-$D\bar D$ decays (pink).
}
\label{pmu_vs_umiss_bes3}
\end{figure}

Figure~\ref{pmu_vs_umiss_bes3}(a) and (b) show the scatter-plots
of the momentum of the identified muon satisfying the requirement
for selecting $D^+\rightarrow \mu^+\nu_{\mu}$ decay versus $M^2_{miss}$,
where the blue box in Fig.~\ref{pmu_vs_umiss_bes3}(a) shows the signal region
for $D^+\rightarrow \mu^+\nu_{\mu}$ decays.
Within the signal region,
there are $425$ candidate events for $D^+ \rightarrow \mu^+\nu_{\mu}$ decay.
The two concentrated clusters out side of the signal region are
from $D^+$ non-leptonic decays and some other background events.
The events whose peak is around 0.25 GeV$^2$/$c^4$ in $M^2_{miss}$ are
mainly from $D^+ \rightarrow K_L^0\pi^+$ decays, where $K_L^0$ is missing.
Projecting the events for which the identified muon momentum being in the range from 0.8 to 1.1 GeV/$c$
onto the horizontal scale yields
the $M^2_{miss}$ distribution as shown in Fig.~\ref{pmu_vs_umiss_bes3}(c),
where the difficultly suppressed backgrounds from
$D^+ \rightarrow K_L^0\pi^+$ decays in CLEO-c measurement~\cite{cleo-c_fD_2008}
are effectively suppressed due to that they use the MUC measurements to identify the muon.

Some non-purely leptonic decay events from the $D^+$, $D^0$,
$\gamma \psi(3686)$, $\gamma J/\psi$,
$\psi(3770) \rightarrow {\rm non}-D\bar D$,
$\tau^+\tau^-$ decays as well as continuum light hadron production
may also satisfy the selection criteria
and are the background events to the purely
leptonic decay events.
These background events must be subtracted off.
The number of the background events can be estimated by analyzing
different kinds of Monte Carlo simulation events.
Detailed Monte Carlo studies show that there are
$47.7 \pm 2.3 \pm 1.3$
background events in $425$ candidates
for $D^+ \rightarrow \mu^+\nu_{\mu}$ decays,
where the first error is due to Monte Carlo statistic and second systematic
arising from uncertainties in the branching fractions
or production cross sections.

After subtracting the number of background events,
$377.3\pm 20.6 \pm 2.6$ signal events
for $D^+ \rightarrow \mu^+\nu_{\mu}$ decay are retained,
where the first error is statistical and the second systematic arising from the uncertainty
of the background estimation.

The overall efficiency for observing the decay ${D^+\rightarrow \mu^+\nu_{\mu}}$
is obtained by analyzing full Monte Carlo simulation events
of $D^+\rightarrow \mu^+\nu_{\mu}$ VS $D^-$ tags
and combining with $\mu^+$ reconstruction efficiency in the MUC.
The $\mu^+$ reconstruction efficiency in the MUC is measured with
muon samples selected from the same data taken at 3.773 GeV. The overall efficiency is
$0.6382~\pm 0.0015$.

With $1565953$ singly tagged $D^-$ mesons, $377.3\pm 20.6 \pm 2.6$
$D^+ \rightarrow \mu^+\nu_{\mu}$ decay events observed and
the efficiency $0.6382~\pm 0.0015$,
the BES-III collaboration obtain the branching fraction
$$B(D^+ \to \mu^+\nu_{\mu})=(3.74 \pm 0.21 \pm 0.06)\times 10^{-4}~~({\rm BESIII~Preliminary}), $$
where the first error is statistical and the second systematic.
This measured branching fraction is consistent within error with
world average of 
$B(D^+ \to \mu^+\nu_{\mu})=(3.82 \pm 0.33)\times 10^{-4}$~\cite{pdg2010},
but with more precision.

The decay constant $f_{D^+}$ can  be obtained
by inserting the measured branching fraction, the mass of the muon,
the mass of the $D^+$ meson, the CKM matrix element
$|V_{\rm cd}|= 0.2252\pm0.0007$ from the CKMFitter~\cite{pdg2010}
$G_F$
and the lifetime of the $D^+$ meson~\cite{pdg2010}
into Eq.(\ref{eq01}), which yields
$$f_{D^+} = (203.91 \pm 5.72 \pm 1.97)~~\rm MeV~~({\rm BESIII~Preliminary}),$$
\noindent
where the first errors are statistical and the second systematic arising
mainly from the uncertainties in
the measured branching fraction ($1.7\%$),
the CKM matrix element $|V_{\rm cd}|$ ($0.3 \%$),
and the lifetime of the $D^+$ meson ($0.7\%$)~\cite{pdg2010}.
The total systematic error is $1.0\%$.

\subsection{Semi-leptonic decays of $D^0$ meson}

Semileptonic decays are an excellent environment for precision 
measurements of the Cabibbo-Kobayashi-Maskawa (CKM) matrix elements. 
However, Extract of the CKM weak parameters require knowledge of 
strong interaction effects. These can be parametrized by form
factors. Techniques such as lattice quantum chromodynamics
offer increasingly precise calculations of these form factors, 
but as the uncertainties in the predictions shrink, experimental 
validation of the results becomes increasingly important.

The $D^0$ mesons are produced from decays $\psi(3770) \to
D^0 \bar{D}^0$. The tagged-$D^0$ is reconstructed from four
hadronic modes. The amount of signal events is determined by 
fitting the distribution of $U_{miss} = E_{miss} - |\vec{p}_{miss}|$.
Based on data sample of 0.92 fb$^{-1}$, preliminary results
on the branching fractions are measured as:
$\BR(\bar{D}^0\to K^+e^-\nu) = (3.542 \pm 0.030 \pm 0.067) \times 10^{-2}$
and
$\BR(\bar{D}^0\to\pi^+e^-\nu) = (0.288 \pm 0.008 \pm 0.005) \times 10^{-2}$.
Note that results are based on approximately one third of 
the statistics and systematic errors are preliminary.
The analysis with the full 2.9 fb$^{-1}$ data and the form factor
measurement are ongoing.

\subsection{$D^0 \to \gamma\gamma$}

In the Standard Model flavor-changing neutral currents
are forbidden at tree level \cite{fcnc}. These decays are allowed
at higher order. To date, measurements of radiative
decays of charm mesons are consistent with results
of theoretical calculations that include both short-distance
and long-distance contributions and predict decay
rates several orders of magnitude below the sensitivity
of current experiments. While these rates are small,
it has been postulated that new physics processes can
lead to significant enhancements \cite{fcnc_enhance}.

The $D^0$ mesons are produced from $\psi(3770) \to D^0 \bar{D}^0$. 
The signal yields for $D^0\to\gamma\gamma$ is obtained by
fitting the distribution $\Delta E = E_{\gamma\gamma} - E_{beam}$. 
No signal is observed. The upper limit of branching fraction is
determined to be $\BR(D^0\gamma\gamma) < 4.6 \times 10^{-6}$ 
(preliminary) at 90\% confidence level. This upper limit is tighter 
than that in PDG~\cite{pdg2010} but not as stringent as measured by 
BaBar Collaboration ($2.4 \times 10^{-6}$) \cite{fcnc_babar}.

\section{Other activities}

\subsection{$\tau$ mass measurement}
As fundamental parameters of the standard model, masses of 
quarks and leptons cannot be determined by the theory and 
must be measured. A precise measurement of the mass of the
$\tau$ lepton is important for testing lepton universality
and for calculating branching fractions that depend on the
$\tau$ mass.
The $\tau$ mass measurement benefits from a high-precision
beam-energy monitor based on the detection of Compton scattering
of back-scattered photons from a high powered single-mode
infrared laser.  This system has been commissioned and
routinely measures the beam energy with a precision
of $\delta_{E_{\rm beam}}/E_{\rm beam} \simeq 10^{-5}$~\cite{bes3_bem}.
A optimized energy scan near the $\tau$ pair production 
threshold has been performed. About 24 pb$^{-1}$ of data, 
distributed over 4 scan points, have been collected. This 
work is based on the combined data from the $ee, e\mu, eh, 
\mu\mu, \mu h, hh, e\rho, \mu\rho$ and $\pi\rho$ final states, 
where $h$ denotes a charged $\pi$ or K. The mass of the 
$\tau$ lepton is measured from a maximum likelihood fit to 
the $\tau$ pair production cross section data which yields 
the value of $m_\tau$ with a precision $<0.3$ MeV.
With 100 pb$^{-1}$ data planned later this year, the
precision will eventually reach to $<0.1$ MeV.

\subsection{Additional $\jpsi$ and $\psip$ data samples.}

Most of the results reported above are based on 106M event $\psip$ 
and 225M event $\jpsi$ data samples. Earlier this year BESIII 
collected another $\sim$0.4B $\psip$ events and $\sim$1.0B $\jpsi$ 
events.  These samples will be used, among other things, 
for detailed PWA of the many  unassigned resonance peaks that 
have been seen, studies of baryon spectroscopy, and 
high-statistics measurements of isospin-violating processes that 
are proving to be valuable probes of the structure of near-threshold
resonances.  In addition, with the huge $\jpsi$ data sample, the
expected SM level for weak decays of the $\jpsi$ to final states 
containing a single $D$ or $D_s$ meson can be accessed and searches 
for non-SM weak decays and lepton-flavor-violating decays, such as 
$\jpsi\rt e^+\mu^-$, will have interesting sensitivity.

\subsection{$R$ measurement and QCD studies}

Before shutdown for summer maintenance this year, BESIII also 
collected data at 4 low energies: 2.23, 2.4, 2.8 and 3.4 GeV.
At each energy point, the number of inclusive hadronic events
is more than $10k$, which will reduce the statistics uncertainty 
for $R$ measurement down to 1\% level, and thus make it possible 
for a $\sim 3\%$ precision measurement.

Other QCD studies, like fragmentation function, baryon form factors, 
multihadron production, are also expected with the data samples.

\section{Concluding remarks and future prospect}

The BESIII experiment at the Institute of High Energy Physics
in Beijing, China is up and running and producing interesting
results on a variety fo topics.  The BEPCII collider is performing
near design levels and the BESIII detector performance is excellent.
We expect to produce many interesting new results in the coming decade.  

BESIII plans to redo the total cross section measurements for 
$\ee\rt hadrons$ with higher precision over the entire accessible 
c.m. energy range, measure $\piz$ and $\eta$ form factors in 
two-photon collisions, remeasure the $\tau$ mass with much 
improved accuracy, and do studies of the recently discovered 
$XYZ$ mesons.
   
Cross section measurement scans will cover c.m. energies from near the
nucleon-antinucleon threshold up to the $\Lambda^+_c \Lambda^-_c$
threshold.  The data near the nucleon-antinucleon threshold will be used 
to measure neutron form factors~\cite{baldini}.  

Data taken in a dedicated run at $E_{c.m.}\simeq 4260$~MeV will
be used to study $Y(4260)$ decays.  Sensitive
searches for possible new, exotic mesons that decay to $\pi^+\jpsi$
and $\pi^+ h_c$, analogous to the $Z_1 (10610)^+$
and $Z_2(10650)^+$ mesons seen by Belle in the $\bbbar$
bottomonium meson system~\cite{belle_Z_b}, will be performed for
$\pipi\jpsi$ and $\pipi h_c$ final states. 

Ultimately, over the next a few years, BESIII intends to collect 
a total of $\sim 10$~fb$^{-1}$ at the $\psi(3770)$ for $D$ meson 
measurements and a comparable sample at higher energy, e.g. 4.17 GeV,
for $D_s$ meson studies.

\section{Aknowledgements}
I appreciate the hard work by the local committee for organizing 
the meeting, especially by Prof. Qun Wang and Prof. Zhengguo Zhao.
I thank my BESIII colleagues for allowing me to represent them and
for generating the results discussed herein. Liaoyuan Dong and
Changzheng Yuan provided materials for preparing my presentation,
and I also benefited from informative conversations with Haibo Li,
Gang Rong, and suggestions from Xiaorui Lu.
The BESIII collaboration thanks the staff of BEPCII and the
computing center for their hard efforts.


\begin{thebibliography}{99}

\bibitem{tau-mass} J.Z.~Bai {\etal} (BES Collaboration),
Phys. Rev. Lett. \textbf{69}, 3021 (1992).

\bibitem{bes_R} J.Z.~Bai {\etal} (BES Collaboration),
Phys. Rev. Lett. \textbf{84}, 594 (2000).

\bibitem{bes_R2} J.Z.~Bai {\etal} (BESII Collaboration),
Phys. Rev. Lett. \textbf{88}, 101802 (2002).

\bibitem{Higgs-mass} H.~Burkhardt and B.~Pietrzyk,
Phys. Lett. \textbf{B513}, 46 (2001).

\bibitem{bes_sigma} M.~Ablikim {\etal} (BESII Collaboration),
Phys. Lett. \textbf{B598}, 149 (2004).

\bibitem{bes_kappa} M.~Ablikim {\etal} (BESII Collaboration),
Phys. Lett. \textbf{B633}, 681 (2006).

\bibitem{bes_x1860} J.Z.~Bai {\etal} (BESII Collaboration),
Phys. Rev. Lett. \textbf{91}, 022001 (2003).

\bibitem{bes3_x1860} M.~Ablikim {\etal} (BESIII Collaboration),
Phys. Rev. Lett. 108, 112003 (2012);
see also M.~Ablikim {\etal} (BESIII Collaboration),
Chin. Phys. C \textbf{34}, 421 (2010).

\bibitem{julich_fsi} E.~Klempt, F.~Bradamente, A.~Martin and J.M.~Richard,
Phys. Rep. \textbf{368}, 119 (2002). 

\bibitem{bes3_eta1405} M.~Ablikim {\etal} (BESIII Collaboration),
Phys.Rev.Lett. 108, 182001 (2012).

\bibitem{zhao} J.J.~Wu, X.-H.~Liu, Q.~Zhao and B.S.~Zou,
Phys. Rev. Lett. \textbf{108}, 081803 (2012). 

\bibitem{pdg2010} K.~Nakamura, {\etal} (Particle Data Group),
Jour. Phys. G \textbf{37}, 075021 (2010).

\bibitem{bes3_rhopi} M.~Ablikim {\etal} (BESIII Collaboration),
Phys.Lett. B710, 594 (2012).

\bibitem{Ablikim:2006}
  M.~Ablikim {\it et al.} [BES Collaboration],
  Phys.\ Rev.\ Lett. {\bf 96}, 162002 (2006).

\bibitem{oneorder}
K$\ddot{o}$pke L, Wermes N. $\jpsi$ Decays. CERN-EP/88-93, Physics Reports 174 (1989) 67-227: CERN,
CH-1211 Geneva 23, Switzerland.

\bibitem{Bing-An:2006}
  B. A. Li,
  Phys.\ Rev.\ D {\bf 74}, 054017 (2006).

\bibitem{Kung-Ta:2006}
 K. T. Chao, hep-ph/0602190.

\bibitem{Bicudo:2007}
  P. Bicudo {\it et al.},
  Eur.\ Phys.\ J.\ C {\bf 52}, 363$-$374 (2007)

\bibitem{Qiang:2006}
  Q. Zhao {\it et al.},
  Phys.\ Rev.\ D {\bf 74}, 114025 (2006).

\bibitem{D.V.:2006}
 D. V. Bugg, hep-ph/0603018.

 \bibitem{belle}
 C. Liu {\it et al.} [Belle Collaboration],
 Phys.\ Rev.\ D {\bf 79}, 0701102(R)(2009).

\bibitem{nora} For a recent review see N.~Brambilla {\etal}, 
Eur. Phys. J. C \textbf{71}, 1534 (2011). 

\bibitem{belle_etacp} S.-K.~Choi {\etal} (Belle Collaboration),
Phys. Rev. Lett. \textbf{89}, 102001 (2002).

\bibitem{cleo_hc} J.~Rosner {\etal} (CLEO Collaboration),
Phys. Rev. Lett. \textbf{95}, 102003 (2005).

\bibitem{bes3_hc1} M.~Ablikim {\etal} (BESIII Collaboration),
Phys. Rev. Lett. \textbf{104}, 132002 (2010).

\bibitem{kuang} Y.P.~Kuang, S.F.~Tuan and T.M.~Yan
Phys. Rev. D \textbf{37}, 1210 (1988), P.~Ko,
Phys. Rev. D \textbf{52}, 1710 (1995),
Y.P.~Kuang, Phys. Rev. D \textbf{65}, 094024 (2002),
and S.~Godfrey and J.~Rosner,
Phys. Rev. D \textbf{66}, 014012 (2002). 
 
\bibitem{bali} See, for example, G.~Bali {\etal}, PoS LATTICE2010, 134 (2011),
arXiv:1011.2195 [hep-lat].

\bibitem{bes3_etac} M.~Ablikim {\etal} (BESIII Collaboration),
Phys.Rev.Lett. 108, 222002 (2012).

\bibitem{cleo_etac} R.~Mitchell {\etal} (CLEO Collaboration),
Phys. Rev. Lett. \textbf{102}, 011801 (2009). 

\bibitem{babar_B2Ketacp} S.-K. Choi {\it et al.}
(Belle Collaboration), Phys. Rev. Lett. {\bf 89}, 102001 (2002).

\bibitem{babar_2gamEtacp} B. Aubert {\it et al.}
(BaBar Collaboration), Phys. Rev. Lett. {\bf 92}, 142002 (2004).

\bibitem{cleo_2gamEtacp} D. M. Asner {\it et al.}
(CLEO Collaboration), Phys. Rev. Lett. {\bf 92}, 142001 (2004).

\bibitem{belle_Jpsiccbar} K.~Abe {\it et al.}
(Belle Collaboration),  Phys.\ Rev.\ Lett.\  {\bf 89}, 142001
(2002).

\bibitem{babar_Jpsiccbar} B. Aubert {\it et al.}
(BaBar Collaboration), Phys. Rev. D {\bf 72}, 031101 (2005).

\bibitem{bes3etacp} M.~Ablikim {\etal} (BESIII Collaboration),
Phys. Rev. Lett. \textbf{109}, 042003 (2012).

\bibitem{Pachucki:1996jw}
  K.~Pachucki {\it et al.},
  J.\ Phys.\ B {\bf 29}, 177 (1996);
  A.~Quattropani and F.~Bassani,
  Phys.\ Rev.\ Lett.\  {\bf 50}, 1258 (1983).

\bibitem{Bai:2004cg}
  J.~Z.~Bai {\it et al.}  [BES Collaboration],
  Phys.\ Rev.\  D {\bf 70}, 012006 (2004);
  M.~Ablikim {\it et al.}  [BES Collaboration],
  Phys.\ Rev.\  D {\bf 71}, 092002 (2005).

\bibitem{Adam:2005uh}
  N.~E.~Adam {\it et al.}  [CLEO Collaboration],
  Phys.\ Rev.\ Lett.\  {\bf 94}, 232002 (2005). 

\bibitem{:2008kb}
  H.~Mendez {\it et al.}  [CLEO Collaboration],
  Phys.\ Rev.\  D {\bf 78}, 011102 (2008).

\bibitem{bes3_psip2ggjpsi} M.~Ablikim {\etal} (BESIII Collaboration),
arXiv:1204.0246.

\bibitem{asner} See, for example, D.M.~Asner and W.M.~Sun,
Phys. Rev. D \textbf{73}, 034024 (2006).

\bibitem{cleo-c_fD_2008} B.I. Eisenstein $et ~al.$ (CLEO Collaboration), 
Phys. Rev. D {\bf 78}, 052003 (2008).

\bibitem{fcnc} S. L. Glashow, J. Iliopoulos, and L. Maiani, 
Phys. Rev. D 2, 1285 (Oct 1970).

\bibitem{fcnc_enhance} S. Prelovsek and D.Wyler, 
Phys. Lett. B 500, 304 (2001), hepph/0012116.

\bibitem{fcnc_babar} J. P. Lees $et ~al.$ (BarBar Collaboration), 
Phys. Rev. D 85, 091107(R) (2012).

\bibitem{bes3_bem} E.V.~Abakumova {\etal},
Nucl. Instrum. Meth. \textbf{A659}, 21 (2011).

\bibitem{baldini} R.~Baldini, S.~Pacetti and A.~Zallo,
Nucl. Phys. Proc. Suppl. \textbf{219-220}, 32 (2011).

\bibitem{belle_Z_b} A.~Bondar {\etal} (Belle Collaboration),
Phys.Rev.Lett. 108, 122001 (2012).

\end{thebibliography}
\end{document}